\documentclass[useAMS,usenatbib]{mnras}
\usepackage{amsmath}
\usepackage{mathtools}
\usepackage{amssymb}
\usepackage{graphicx}
\usepackage{multirow}
\usepackage{color}
\usepackage{array}
\usepackage{graphicx}

\title[Predicting BH mass using neural networks]{Predicting the black hole mass and correlations in X-ray reverberating AGN using neural networks}

\author[P. Chainakun et al.]{P. Chainakun$^{1,2}$\thanks{E-mail: \href{mailto:pchainakun@g.sut.ac.th}{pchainakun@g.sut.ac.th}}, I. Fongkaew$^{1}$, S. Hancock$^{3}$, A. J. Young$^{3}$\\
$^1$School of Physics, Institute of Science, Suranaree University of Technology, Nakhon Ratchasima 30000, Thailand\\
$^2$Centre of Excellence in High Energy Physics and Astrophysics, Suranaree University of Technology, Nakhon Ratchasima 30000, Thailand\\
$^3$HH Wills Physics Laboratory, Tyndall Avenue, Bristol BS8 1TL, UK}
\date{Accepted XXX. Received YYY; in original form ZZZ}

\pubyear{2021}

\begin{document}
\label{firstpage}
\pagerange{\pageref{firstpage}--\pageref{lastpage}}
\maketitle

\begin{abstract}

We develop neural network models to predict the black hole mass using 22 reverberating AGN samples in the {\it XMM-Newton} archive. The model features include the fractional excess variance ($F_{\rm var}$) in 2--10 keV band, Fe-K lag amplitude, 2--10 keV photon counts and redshift. We find that the prediction accuracy of the neural network model is significantly higher than what is obtained from the traditional linear regression method. Our predicted mass can be confined within $\pm (2$--5) per cent of the true value, suggesting that the neural network technique is a promising and independent way to constrain the black hole mass. We also apply the model to 21 non-reverberating AGN to rule out their possibility to exhibit the lags (some have too small mass and $F_{\rm var}$, while some have too large mass and $F_{\rm var}$ that contradict the $F_{\rm var}$--lag--mass relation in reverberating AGN). We also simulate 3200 reverberating AGN samples using the multi-feature parameter space from the neural network model to investigate the global relations if the number of reverberating AGN increases. We find that the $F_{\rm var}$--mass anti-correlation is likely stronger with increasing number of newly-discovered reverberating AGN. Contrarily, to maintain the lag--mass scaling relation, the tight anti-correlation between the lag and $F_{\rm var}$ must preserve. In an extreme case, the lag--mass correlation coefficient can significantly decrease and, if observed, may suggest the extended corona framework where their observed lags are more driven by the coronal property rather than geometry.  


\end{abstract}


\section{Introduction}

The X-ray variability has become a powerful tool to probe the inner accretion flow around the central black holes in both active galactic nuclei (AGN) and X-ray binaries. The X-ray reverberation mapping is one of the timing analysis techniques that measures the time delays associated with the light-travel time between the X-ray photons from the direct coronal emission and back-scattered photons from the accretion disc \citep[see][for a review]{Uttley2014, Cackett2021}. Due to a longer distance travelled by the reflected photons, the reflection-dominated bands (the soft excess, Fe-K and Compton hump bands) lag behind the continuum-dominated bands \citep[e.g.][]{Fabian2009, Kara2013a, Kara2013c, Kara2014, Zoghbi2014, Kara2019, Alston2020, Vincentelli2020}. The amplitude of the lag depends on the geometry of the source, so providing us a tool to constrain the disc-corona geometry. 

\cite{Demarco2013} performed a systematic look at
the frequency-dependent time lags between the soft ($\sim 0.3–1$ keV) and hard ($\sim 1–4$ keV) bands in AGN and found that the amplitude of the soft lag scales with the black hole mass. \cite{Kara2016} performed a systematic study of the X-ray reverberation lags of all Seyfert galaxies available in the \emph{XMM–Newton} archive and reported a number of sources that exhibited Fe-K reverberation lags. They confirmed the lag-mass scaling relation and found the correlation between the height of the corona and the mass accretion rate of these reverberating AGN. \cite{King2017} found that the radio Eddington luminosity inversely correlates with the X-ray reflection fraction, and positively scales with the path length between the X-ray source and the accretion disc. Modelling the X-ray reverberation lags to map the extreme region near a supermassive black hole has been carried out intensively using both lamppost geometry \citep{Wilkins2013, Cackett2014, Emmanoulopoulos2014, Chainakun2015, Chainakun2016, Epitropakis2016, Caballero2018, Ingram2019} and extended corona model \citep{Wilkins2016,Chainakun2017,Chainakun2019b}. The X-ray reverberation technique has already been applied to various scenarios such as the photon reflection off the accretion flow in the tidal disruption event \citep{Kara2016b}, the reflection from different hot-flow zones in X-ray binaries \citep{Mahmoud2019, Chainakun2021b, Kawamura2021} and the multiple scattering from the disc wind in ultraluminous X-ray sources \citep{Luangtip2021}. 

Furthermore, \cite{Alston2020} suggested that the height of the X-ray corona in IRAS13224--3809 increased with the source luminosity. This is also supported by \cite{Caballero2020} who found significant variations in the X-ray source height from $\sim 3-5 r_{\rm g}$ when the X-ray luminosity is $\sim 1.5-3$ per cent of the Eddington limit, to $\sim 10 r_{\rm g}$ when the luminosity doubles. Recently, Hancock et al. (in prep) investigated the time-average and lag-frequency spectra of 20 AGN covering 121 \emph{XMM-Newton} observations and separated them into 3--4 groups of similarly observed spectral states. The reflection fraction was found to be strongly correlated to the power-law photon index that suggested dynamics of the emitting region.

The X-ray reverberation features can be imprinted in the profiles of the power spectral density (PSD) that describes the variability power on different timescales. The oscillatory structures seen in the PSD of AGN can be interpreted as the reverberation signatures that relate to the geometry of the system such as the coronal height and the inclination \citep{Papadakis2016,Emmanoulopoulos2016, Chainakun2019a}. In \cite{Chainakun2021}, we developed machine learning (ML) models, based on dictionary learning and support vector machine algorithms, to extract the X-ray reverberation signatures on the PSD profiles of AGN, and used them to predict the coronal height. The variability amplitude in light curves can be estimated using the fractional excess variance ($F_{\rm var}$), which can be constructed by integrating the PSD between two frequencies, or from the mean and the rms amplitude of the light curve \citep[e.g.][]{Vaughan2003}. Recently, the $F_{\rm var}$ spectra have been used to probe the intrinsic and environmental absorption origins for the X-ray variability in AGN \citep{Parker2021}. 

The potential use of the ML techniques in the X-ray reverberation analysis has been elaborated and discussed in \cite{Chainakun2021}. Here, we employ the key information of the X-ray reverberating AGN to develop a neural network model in order to predict the black hole mass. We consider the fundamental parameters that can be derived from the X-ray observations. These parameters include the Fe-K lag amplitude, the $F_{\rm var}$, the photon counts ($C$), the bolometric luminosity ($L_{\rm bol}$) and the redshift ($z$). The source height and the reflection fraction are not considered because their values are dependent on the assumed geometry and the choice of the reflection models.

The ultimate goal is to test how well the neural network model, trained using only fundamental X-ray observational parameters, can predict the central mass. The prediction accuracy obtained from the neural network model is reported and compared to what is obtained from the standard linear regression model. After that, we plot the mass distribution against the model features such as the $F_{\rm var}$ and the lag amplitude to investigate the global relations between them all simultaneously. From the parameter space constrained by the model, we can simulate more samples of reverberating AGN and investigate the correlations between their parameters. It provides hints of how the known-existing correlations will change if more reverberating AGN are discovered. The AGN data used for training the machine are presented in Section 2. The neural network algorithm and the development of the ML models to predict the black hole mass are explained in Section 3. We evaluate the models and present the results in Section 4. We discuss the optimization results and the obtained correlations in Section 5, while the conclusion is provided in Section 6.

\section{AGN data}
We use the \emph{XMM-Newton} data previously reported and analyzed by \cite{Kara2016} where the selected samples have $\gtrsim 40$ ks exposure time and show some variability. However, we select only the data that display the Fe-K reverberation features in the lag spectra whose lower and upper bounds of the lags can be constrained. Based on these criteria, there are 22 AGN sources in total. These AGN samples and their parameter values are listed in Table~\ref{tab_data}. The Fe-K lag amplitude, $F_{\rm var}$, log($L_{\rm bol}$) and log($C$) of each AGN are average values from those of all available observations that fit the criteria. $F_{\rm var}=\sigma_{\rm rms}/\bar{x}$ is the fractional excess variance of the data in 2--10 keV band where $\sigma_{\rm rms}$ is the rms amplitude and $\bar{x}$ is the mean count rates \citep{Vaughan2003}. $C$ is the photon counts in the 2--10 keV band. $L_{\rm bol}$ is the mean bolometric luminosities calculated from the SEDs \citep{Wang2004, Vasudevan2007, Vasudevan2009, Vasudevan2010}. The masses that are estimated by the optical reverberation techniques are from the public web data base \citep{Bentz2015} using the $<f>$ = 4.3 \citep{Grier2013}, similar to \cite{Kara2016}.  

There are also $\sim 21$ AGN samples left, that have $\gtrsim 40$ ks exposure time, show some variability but do not exhibit Fe-K reverberation features on the lag spectra. We spare them in a new data set (Table~\ref{tab_predicted_M}) for further analysis once the neural network model is obtained. In other words, while the reverberating AGN data in Table~\ref{tab_data} are used during the training phase, the non-reverberating AGN data in Table~\ref{tab_predicted_M} are kept unseen, completely new to the machine and are used only for the final evaluation of the model. Note that there are some inconsistencies in the report of the Fe-K response among previous literature. For example, \cite{Wilkins2021} recently found that there was a Fe-K reverberation lag caused by a flare in IZw1, while this source is still included in the non-reverberation sample here (Table 4). We, however, select to follow the standard analysis of \cite{Kara2016} and discuss these inconsistencies later in the Discussion section. 

\begin{table*}
\begin{center}
 \caption{Observed reverberating AGN data used for training and testing the neural network model. The table includes the AGN name, black hole mass, 2-10 keV fractional excess variance, Fe-K lag amplitude, bolometric luminosities, total 2–10 keV counts and redshift. These AGN are all Fe-K reverberating AGN probed by \emph{XMM-Newton} and were previously analyzed by \protect\cite{Kara2016}. The numbers in brackets denote the references where: (1) \protect\cite{Bian2003}; (2) \protect\cite{Ponti2012}; (3) \protect\cite{Agis2014}; (4) \protect\cite{Gonzalez2012}; (5) \protect\cite{Alston2015}; (6) \protect\cite{Schulz1994}; (7) \protect\cite{Marconi2008}; (8) \protect\cite{Alston2014}; (9) \protect\cite{Malizia2008}; (10) \protect\cite{Kara2013a}; (11) \protect\cite{Kara2013c}; (12) \protect\cite{Kara2013b}; (13) \protect\cite{Zoghbi2013}; (14) \protect\cite{Kara2015}; (15) \protect\cite{Zoghbi2012}; (16) \protect\cite{Kara2014}; (17) \protect\cite{Marinucci2014}. (R) indicates the optical reverberation mass estimate. (K) refers to \protect\cite{Kara2016}.}
 \label{tab_data}
\begin{tabular}{lllllll}
\hline
AGN name & log($M/M_{\odot}$) & $F_{\rm var}$ & Lag amplitude (s) & log($L_{\rm bol}$) & log($C$) & z \\
\hline
1H 0707--495 & 6.31 (1) & 0.527 & $47 \pm 16$ (10)& 44.43 & 5.158 & 0.0406 \\
Ark 564  & 6.27 (2) & 0.213 & $92 \pm 65$ (11)& 44.36 & 6.207& 0.0247 \\
ESO 362--G18 & 7.65 (3) & 0.131  & $1562 \pm 606$ (K)& 44.11 & 4.849 & 0.0124 \\
IC 4329A & 8.3 (4) & 0.028 & $696 \pm 331$ (K)& 44.92 & 6.212& 0.0161 \\
IRAS 13224--3809 & 6.8 (4) & 0.612 & $299 \pm 135$ (12)& 45.74 & 4.628 & 0.0658 \\
IRAS 17020+4544 & 6.54 (2) & 0.156 & $128 \pm 88$ (K)& 44.74 & 5.164 & 0.0604 \\
MCG--5--23--16 & 7.92 (5) & 0.074 & $1037 \pm 455$ (13)& 44.30 & 6.681 & 0.0085 \\
Mrk 335	 & 7.23 (R) & 0.177 & $193 \pm 98$ (11)& 45.10 & 5.631 & 0.0258 \\
MS 22549--3712	 & 7.0 (5) & 0.100 & $1500 \pm 850$ (5)& 45.09 & 5.000 & 0.0390 \\
NGC 1365	 & 7.6 (4) & 0.234 & $500 \pm 120$ (14)& 43.99 & 5.940 & 0.0055 \\
NGC 3783	 & 7.371 (R) & 0.066 & $172 \pm 62$ (K) & 44.28 & 6.072 & 0.0097 \\
NGC 4051	 & 6.13 (R) & 0.400 & $90 \pm 30$ (K)& 43.26 & 5.301 & 0.0023 \\
NGC 4151	 & 7.65 (R) & 0.077 & $880 \pm 360$ (15)& 44.01 & 5.899 & 0.0033 \\
NGC 5506	 & 7.4 (4) & 0.097 & $398 \pm 252$ (K) & 44.22 & 6.307 & 0.0062 \\
NGC 5548	 & 7.718 (R) & 0.039 & $311 \pm 109$ (K)& 44.79 & 5.700 & 0.0172 \\
NGC 6860	 & 7.6 (4) & 0.070 & $398 \pm 252$ (K)& 43.71 & 5.530 & 0.0149 \\
NGC 7314	 & 6.7 (6) & 0.223 & $77 \pm 31$ (13)& 42.98 & 5.937 & 0.0048 \\
NGC 7469	 & 6.956 (R) & 0.078 & $1848 \pm 1451$ (K) & 45.10 & 5.806 & 0.0163 \\
PG 1211+143	 & 7.61 (2) & 0.118 & $1179 \pm 980$ (K)& 46.17 & 4.964 & 0.0809 \\
PG 1244+026	 & 7.26 (7) & 0.190 & $726 \pm 306$ (16)& 44.62 & 4.757 & 0.0482 \\
REJ 1034+398	 & 6.6 (8) & 0.170 & $450 \pm 200$ (K)& 44.52 & 5.602 & 0.0424 \\ 
SWIFT J2127.4+5654	 & 7.18 (9) & 0.137 & $408 \pm 127$ (17)& 44.55 & 6.170 & 0.0144 \\

\hline

\end{tabular}
\end{center}
\end{table*}
\nopagebreak

\section{Methods}

\begin{figure}
    \centerline{
        \includegraphics[width=0.5\textwidth]{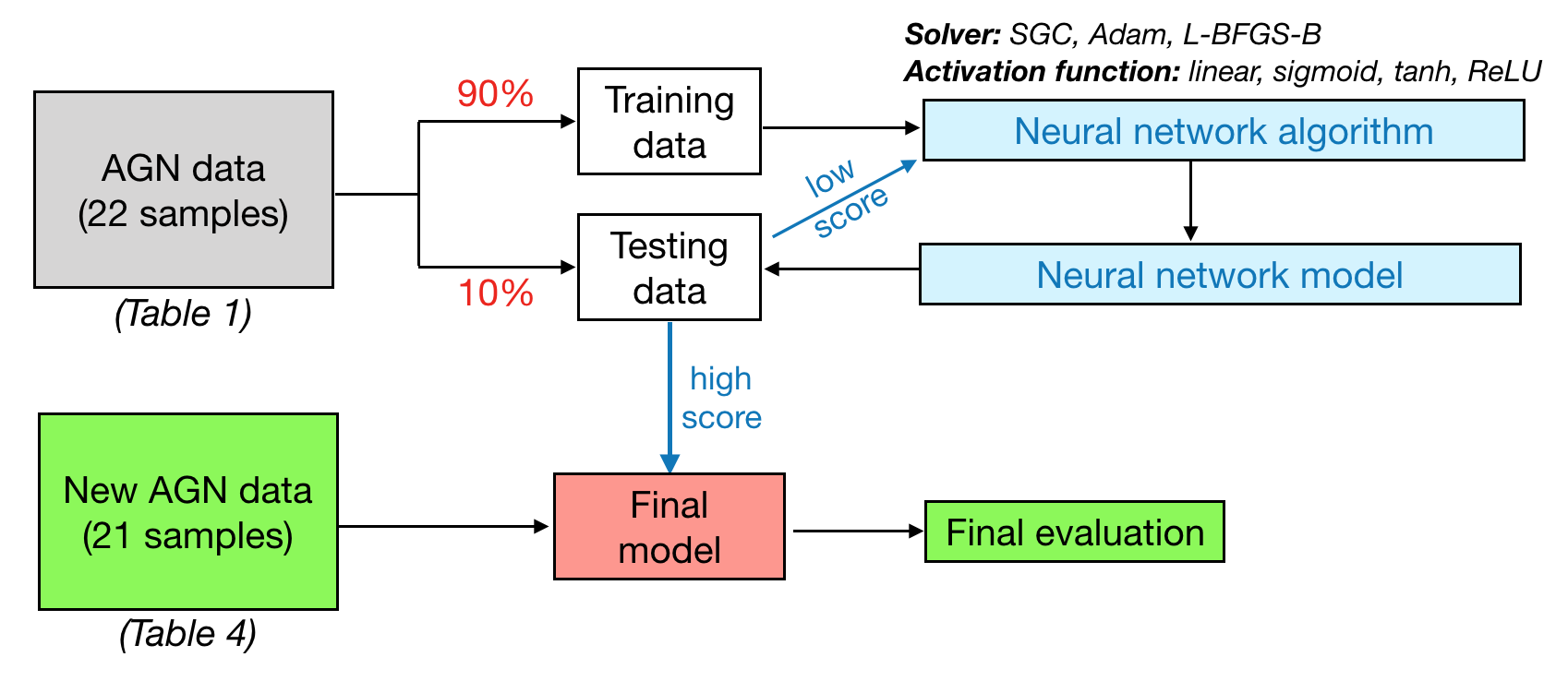}
    }
    \caption{Flowchart of our ML algorithm. The AGN data for training and testing (Table~\ref{tab_data}) are split into the training set (90 per cent) and test set (10 per cent). The neural network algorithm is used to train the machine to make an accurate prediction of the black hole mass. The new AGN data set (Table~\ref{tab_predicted_M}) is also spared for the final evaluation of the model.}
    \label{fig-flowchart}
\end{figure}

We train the machine using the neural network technique so that it can make an accurate prediction of the black hole mass. The flowchart illustrating the training and testing process is presented in Fig.~\ref{fig-flowchart}. The following subsections outline step-by-step the methodology used to explore the nature of the data as well as to develop and evaluate our neural network models. 

\subsection{Pre-processing the data}
The data set used during the training and testing phase contains 22 AGN sources in total, each of which consists of five features which are the lag amplitude, $F_{\rm var}$, log($L_{\rm bol}$), log($C$) and $z$. We train the machine by using only the mean of the data (i.e. ignore the uncertainty from measurements), so that the model becomes as simple as possible. Since the amount of reverberating AGN sources is small, we split them randomly with the sample ratio of 90/10 per cent for training/testing. This can be done using {\tt sklearn.model\_selection.train\_test\_split()} available in {\sc scikit-learn}\footnote{\url{https://scikit-learn.org/}} \citep{Pedregosa2012}.

We also carry out the test when the data for each feature are scaled to be between 0 and 1. This aims to inspect how much the scaling factor affects the performance of the model.

\subsection{Correlations of the data}
We analyze the Pearson and Spearman's rank correlation coefficient ($r_{p}$ and $r_{s}$, respectively) between each model feature and the black hole mass. $r_{p}$ measures the linear correlation between two features resulting in a value between $-1$ and 1. $r_{p} = 1$ and $-1$ mean perfect linear correlation and anti-correlation, respectively. On the other hand, $r_{s}$ assesses the strength and direction of monotonic relationships between two features. Also, $r_{s}$ can be between 1 and $-1$ (between perfect monotonic correlation and anti-correlation, respectively). The features with a very weak correlation with mass are discarded in order to save the computational time during the training phase. Due to a small amount of data, it is also better to keep the number of important features to be as small as possible to avoid data overfitting.

Throughout the paper, the trend of correlations is illustrated using the Sieve diagram that represents the relationship between the categorical variables using observed and expected frequencies under independence. It displays the structure of the data and the model (i.e. the pattern of association) that helps us visualise the relationship between the observed and expected frequencies between two variables. The data analysis and visualization is carried out using the Orange platform which is the data mining toolbox in Python \citep{Demsar2013}.  

\subsection{Neural network algorithm}
The neural network algorithm which is employed here is {\tt sklearn.neural\_network.MLPRegressor()} available in {\sc scikit-learn} \citep{Pedregosa2012}. The {\tt MLPRegressor()} is the Multi-layer Perceptron (MLP) regressor that can optimize, iteratively, the partial derivatives of the loss function at each time step based on the choice of the activation function and solver. The MLP architecture contains a series of layers that consists of neurons and their connections, building up a neural network. The basic unit of a neural network is a neuron that takes inputs, re-processes them and produces one output. The data with $n$ features (i.e. $n$ inputs) can be written as $\textbf{x} = [x_{1}, x_{2}, x_{3}, ..., x_{n}]$. Within a neuron, each input is multiplied by a weight $\textbf{w} = [w_{1}, w_{2}, w_{3}, ..., w_{n}]$. Then, weighted inputs are added together with one bias $b$ and are passed through an activation function $f$ to obtain one output $y$:     
\begin{equation}
y = f(\textbf{x}\cdot \textbf{w} + b) \;.
    \label{eq:nn}
\end{equation} 
The choice of the activation functions depends on the nature of the data. An example function commonly used is the sigmoid function that outputs only numbers between $(0,1)$.  

A neural network is produced by combining many neurons. It can have any number of layers with any number of neurons in those layers. Any layers between the input (first) layer and output (last) layer are referred to as the hidden layer. The appropriate hidden layer size and number of neurons in each layer can be fine-tuned during the training phase.

Performance of the neural network model can be evaluated using either the mean absolute error (MAE) or the mean squared error (MSE):
\begin{equation}
{\rm MAE} = \frac{1}{N} \sum_{i=1}^{N} |y_{i,{\rm true}} - y_{ i, \rm{pred}}| \;,
    \label{eq:mae}
\end{equation}
\begin{equation}
{\rm MSE} = \frac{1}{N} \sum_{i=1}^{N} |y_{i,{\rm true}} - y_{ i, \rm{pred}}|^2 \;,
    \label{eq:mse}
\end{equation}
where $N$ is the number of AGN samples. $y_{\rm true}$ and $y_{\rm pred}$ are the true and predicted values of the black hole mass, respectively. During the training and testing phase, the error loss is estimated in the form of the loss function, $L$, by taking the average overall obtained error. $L$ can then be written as a multivariable function of $L(w_{1}, w_{2}, w_{3}, ..., b_{1}, b_{2}, b_{3},...)$. The change of loss when changing one of these variables (e.g. $\frac{\partial L}{\partial w_{1}}$) can be calculated. Therefore, the goal of training a neural network is to minimize its loss in predicting the black hole mass by finding appropriate weights and biases.

\subsection{Hyperparameter optimization}
Hyperparameters are the parameters that cannot be directly learned by the machine. Fine-tuning them is one of the important processes to improve the performance of the ML model. There are two key hyperparameters for the neural network algorithm which are the hidden layer size and the number of neurons in each layer. Since the number of our inputs is quite small, we begin by setting the number of hidden layers to be 1 and allow the number of neurons to be varied between 1 and 250 neurons. We increase the number of hidden layers only if the prediction accuracy using 1 layer is low.  The learning rate is also a tuning parameter in an optimization algorithm that determines the step size at each iteration while moving toward a minimum of a loss function. An adaptive learning rate algorithm is used by starting from an initial value of 0.001.

Moreover, we investigate three solvers including the standard Stochastic Gradient Descent (SGD), Adam and L-BFGS-B algorithms. Adam is an adaptive algorithm for the first-order gradient-based optimization, which is an extension to SGD. L-BFGS-B is an extension of the Limited-memory Broyden-Fletcher-Goldfarb-Shanno (L-BFGS) algorithm which is a type of second-order optimization algorithm that uses a limited amount of computer memory and can handle bound constraints on variables. These different solvers can provide different optimization results. 

We investigate four activation functions including the identity (linear), logistic (sigmoid), hyperbolic tangent (tanh) and Rectified Linear Unit (ReLU) functions. The activation function helps the network learn complex data so the model can provide accurate predictions. It is clear from eq.~\ref{eq:nn} that a neural network without an activation function, $f$, reduces to a linear regression model. A neural network with the identity activation function is also consistent with the linear regression model. If the neural network model prefers the identity activation function, it means that the model behaves like a single layer network even when we increase the number of layers to our network. This is because summing up the additional layers will still output another linear function which does not further improve the model. 

\section{Results and analysis}

\subsection{Current observed AGN data}
\begin{figure}
    \centerline{
        \includegraphics[width=0.45\textwidth]{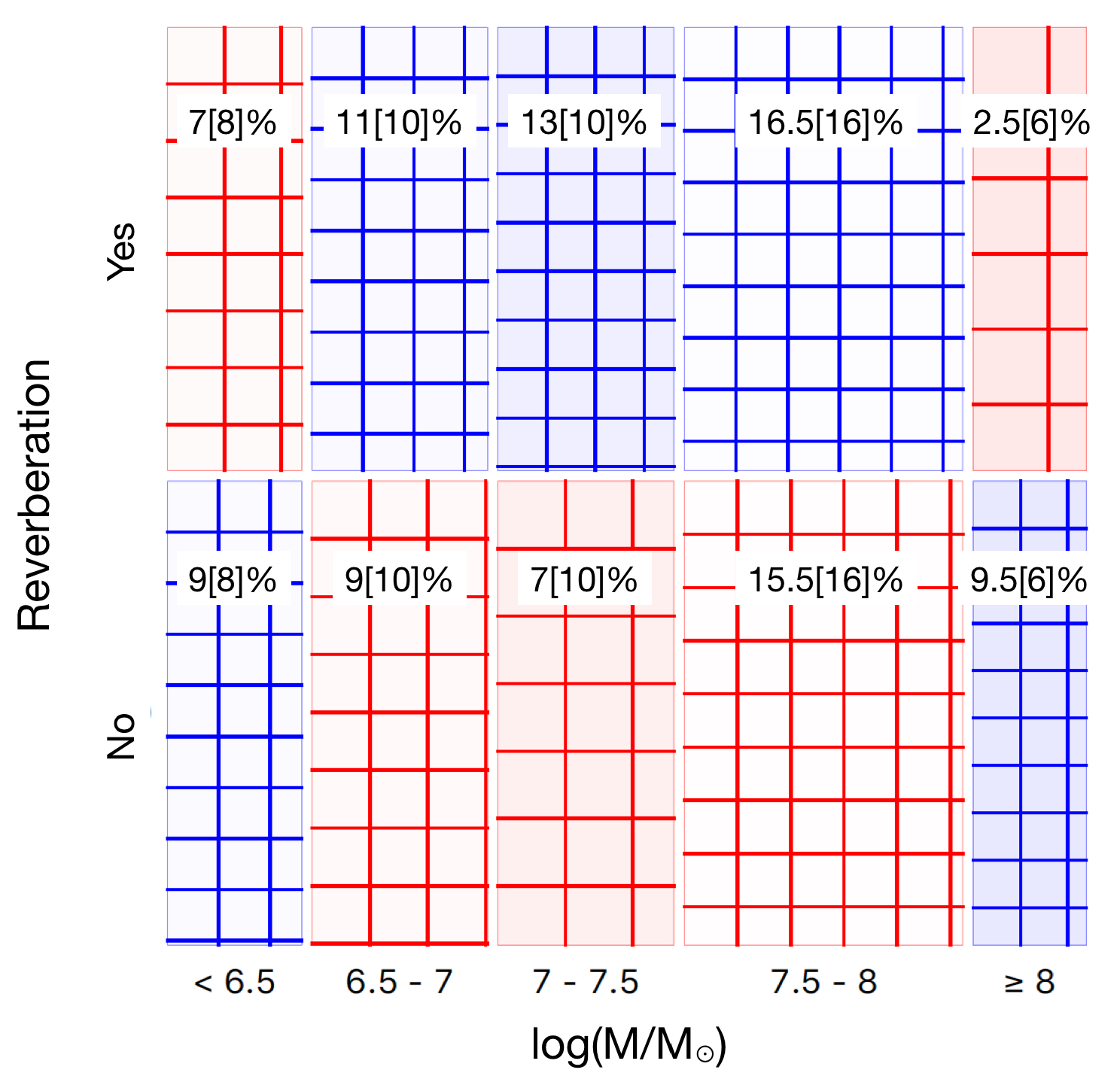}
    }
    \caption{Sieve diagram visualizing frequencies in discovering reverberating AGN (Table~\ref{tab_data}) and non-reverberating AGN (Table~\ref{tab_predicted_M}) across all masses. The area of each rectangular and the number of squares inside are proportional to the expected frequency under independence and the observed frequency, respectively. The density of shading shows the differences between the observed and expected frequency, appearing as blue and red for the positive and negative deviation, respectively. The observed frequencies are shown as percentages while the values in square brackets indicate the per cent of the expected frequencies.}
    \label{fig-reverb-nonreverb}
\end{figure}

First of all, we discretize the data into two groups which are reverberating and non-reverberating AGN samples and explore their distribution in the parameter space. Note that the non-reverberating AGN referred to the sources that the Fe-K reverberation lags cannot be robustly detected with \emph{XMM-Newton} despite that they show some variability. The inconsistency in detecting the Fe-K reverberation in some sources as reported in previous literature is discussed in the Discussion section. Fig.~\ref{fig-reverb-nonreverb} shows the Sieve diagram of the frequencies in discovering these reverberating and non-reverberating AGN regarding to their masses. The Sieve diagram compares the observed frequencies to expected frequencies under the assumption of independence. The expected frequency is proportional to the size of each rectangle (or cell), where the number of squares in each cell represents the observed frequency. Cells whose observed frequency is greater (smaller) than the expected frequency are shown in blue (red), appearing more (less) intense for more (less) deviation. The result suggests a high probability of finding the reverberating AGN if the central mass is in the range of $\sim 10^{6.5}$--$10^{8} M_{\odot}$ (blue shade in the upper row). On the other hand, the samples are dominated by non-reverberating AGN if the central mass is beyond the lower and upper bounds of that range. Especially when the mass is $\gtrsim 10^8 M_{\odot}$, there is a small probability of finding AGN displaying reverberation lags.

\begin{figure}
    \centerline{
        \includegraphics[width=0.45\textwidth]{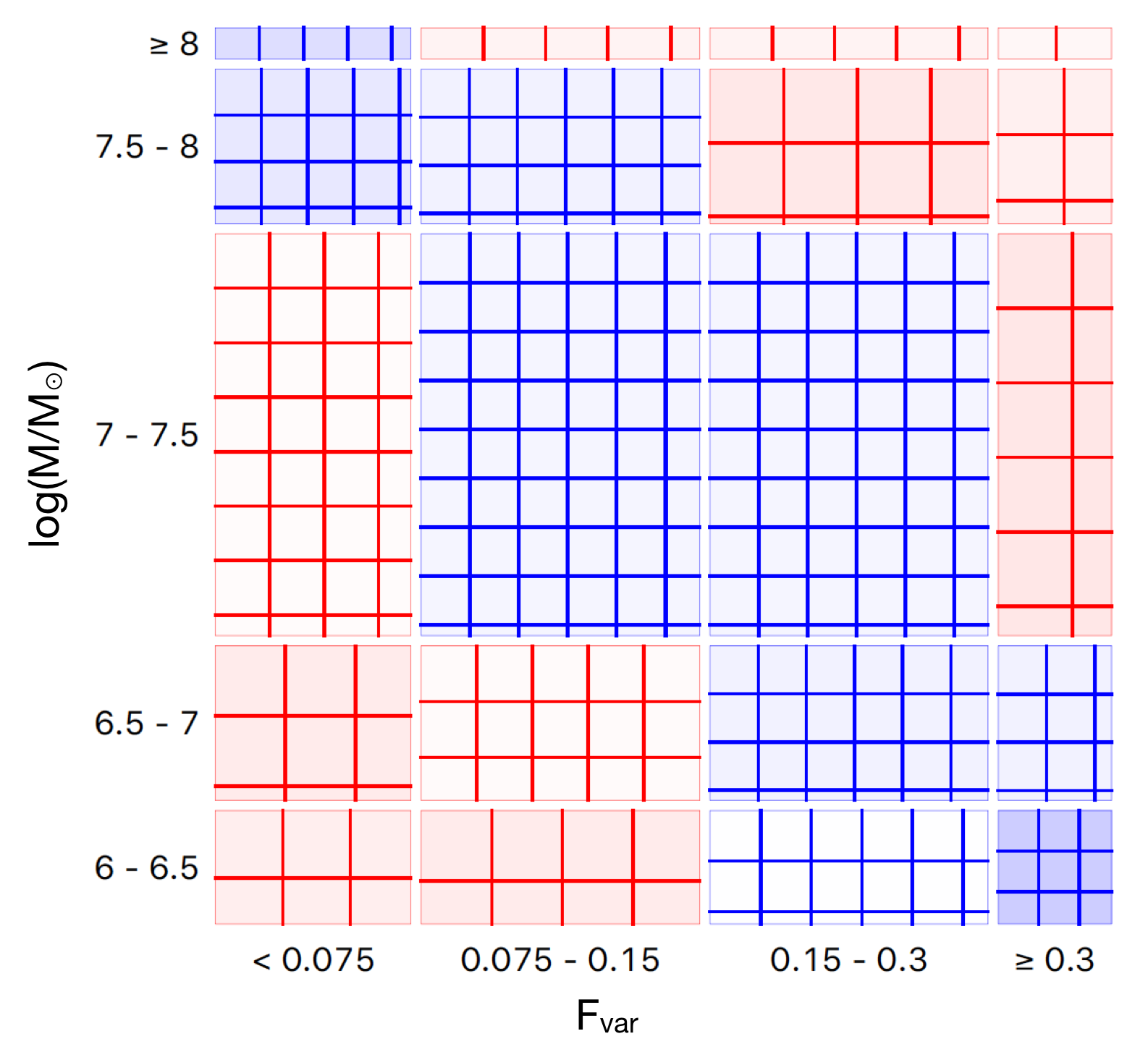}
    }
     \vspace{0.2cm}
    \centerline{
        \includegraphics[width=0.45\textwidth]{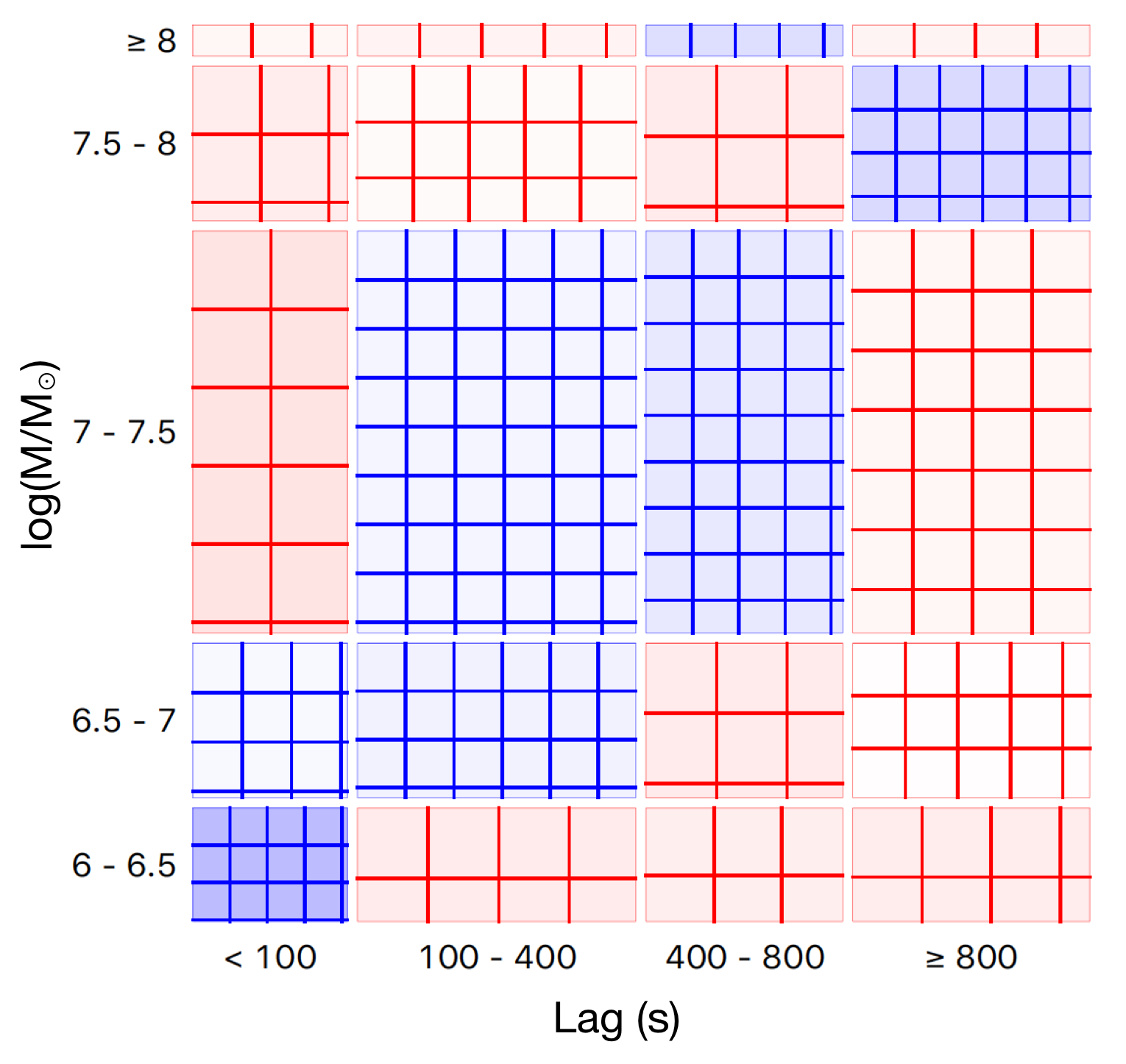}
    }
    \caption{Sieve diagrams for the black hole mass in 22 reverberating AGN samples, dependent on $F_{\rm var}$ (top panel) and lags (bottom panel).}
    \label{fig-s}
\end{figure}

\begin{figure}
    \centerline{
        \includegraphics[width=0.5\textwidth]{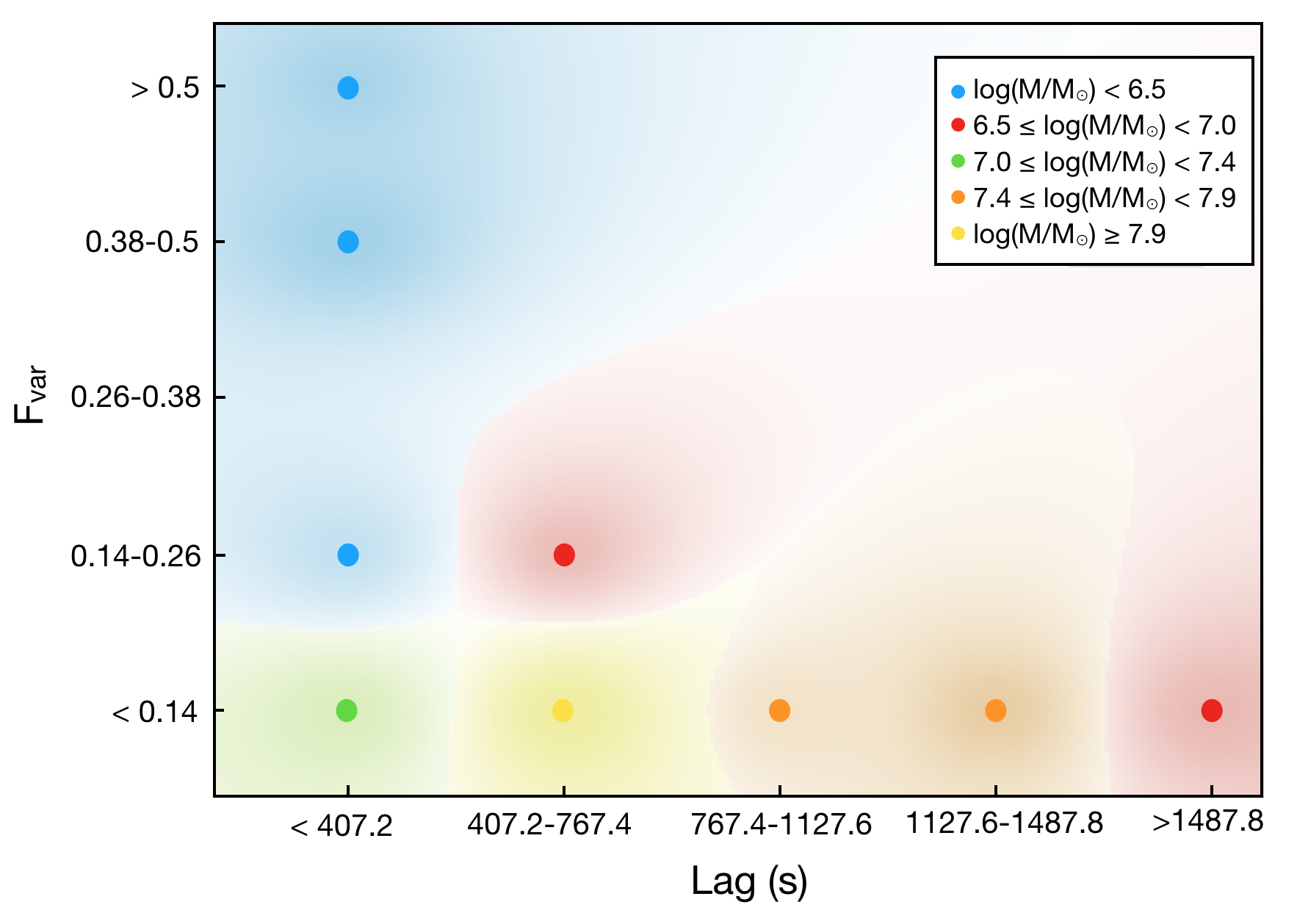}
    }
    \caption{Plot of the mass distribution dependent on the $F_{\rm var}$ and Fe-K lag amplitude of all reverberating AGN in Table~\ref{tab_data}. The data are discretized into 5 groups based on their $F_{\rm var}$, lag amplitude and central mass. Each dot represents the clump of data falling into that particular range of parameters. Different colours represent different corresponding masses as specified in the figure.}
    \label{fig-data-map}
\end{figure}

Now let us focus on the reverberating AGN (22 samples in Table~\ref{tab_data}). The Sieve diagrams representing the chance in finding the Fe-K reverberating AGN having mass, $F_{\rm var}$ and lag amplitude in a specific range of model parameters are shown in Fig.~\ref{fig-s}. The pattern of association is clearly revealed. The number of actual, observed data respondents with the AGN exhibiting larger $F_{\rm var}$ while having smaller mass becomes more than what is expected under the assumption of independence. The known lag-mass scaling relation \citep{Demarco2013, Kara2016} is also suggested. To investigate more on the nature of mass distribution among these reverberating AGN, we discretize the samples into 5 groups based on their $F_{\rm var}$, lags and mass. The result is shown in Fig.~\ref{fig-data-map}. The sources are clumped around the bottom-left portion of the lag--$F_{\rm var}$ parameter space. We can see the trend of decreasing mass with increasing $F_{\rm var}$, and decreasing lag with increasing $F_{\rm var}$. The yellow clump (highest average-mass group) lies in the range of 407--764~s time lags and $F_{\rm var} <0.14$, which is driven by the AGN IC 4329A in particular.

Correlation coefficients between the black hole mass and each model feature are summarized in Table~\ref{tab_corr}. Undoubtedly, $F_{\rm var}$ shows a strong anti-correlation with the black hole mass in both interval and ordinal scales ($r_{p} = -0.636$ and $r_{s}=-0.718$). The lag amplitude also scales with the black hole mass with $r_{s}=0.589$, as expected. Contrarily, log($L_{\rm bol}$) shows a very weak correlation with the mass ($r_{s} = 0.023$). The very weak correlation between log($L_{\rm bol}$) and mass is possibly due to the measurement of $L_{\rm bol}$ itself which is likely model dependent and is easily either overestimated or underestimated based on the inferred bolometric correction factors. We omit this feature to reduce time consumption during the machine training processes. This also helps prevent the model becoming too complex and too specific to the training data. Therefore, there are four features left in our consideration for developing ML models, which are $F_{\rm var}$, lag amplitude, log($C$) and $z$.

\begin{table}
\begin{center}
   \caption{Pearson correlation coefficient, $r_{p}$, and Spearman's rank correlation coefficient, $r_{s}$, between log($M/M_{\odot}$) and each model feature for 22 reverberating AGN samples in the {\it XMM-Newton} archive.} \label{tab_corr}
    \begin{tabular}{lcc}
    \hline
    Features & $r_{p}$ & $r_{s}$  \\
    \hline
     $F_{\rm var}$     &  $-0.636$ & $-0.718$ \\
     Lag amplitude     & $+0.408$ & $+0.589$ \\
     log($L_{\rm bol}$)     &  $+0.172$ & $+0.023$ \\
     log($C$)     &  $+0.295$ & $+0.260$ \\
     $z$     &  $-0.237$ & $-0.233$ \\
    \hline
    \\
     \end{tabular}
\end{center}
\end{table}

\subsection{Mass prediction by the model}   
The best neural network models obtained with different combinations of the features are presented in Table~\ref{tab_ML_model}. Predictions made by the neural networks are much more accurate than what is obtained traditionally from the linear regression method. Note that we fix the number of hidden layers to be 1 and fine-tune the number of neurons with different solvers and activation functions. More features require more neurons because the data are more complex. We find that the best models in all cases prefer the L-BFGS-B solver. The best activation function associated with our data is tanh, except when only the $F_{\rm var}$ is the model feature. The models can make a good prediction of the mass using the $F_{\rm var}$ and the lag amplitude alone, with the $R^2$ of 0.6459 and 0.7302, respectively. The accuracy significantly increases if both $F_{\rm var}$ and lag amplitude are used as the features ($R^2 = 0.9124$).

\begin{table*}
\begin{center}
   \caption{The best neural network models for different combinations of the model features. The best solver, activation function, number of hidden layers (fixed at 1) and number of neurons in each layer are presented in the second, third, fourth and fifth columns, respectively. The corresponding $R^2$ values and the MAE are shown in the sixth and seventh columns. The values in square brackets indicate the corresponding errors obtained from the linear regression model.} \label{tab_ML_model}
    \begin{tabular}{lcccccc}
    \hline
    Features & Solver & Activation & Number & Number & $R^2$ & MAE \\
     & & function  & of layers  & of neurons  & & \\
    
    \hline
     $F_{\rm var}$  & L-BFGS-B & ReLU & 1 & 91 & 0.6459 [0.3558] & 0.2604 [0.4160]   \\
     Lag amplitude & L-BFGS-B & tanh & 1 & 128 & 0.7302 [0.4529]  & 0.2103 [0.1545]  \\
     $F_{\rm var}$ $\oplus$ Lag amplitude  & L-BFGS-B & tanh & 1 & 168 & 0.9124 [0.3512] & 0.1077 [0.4278]   \\
     $F_{\rm var}$ $\oplus$ Lag amplitude $\oplus$ $z$  & L-BFGS-B & tanh & 1 & 164 & 0.9637 [0.3518] & 0.0719 [0.4283]\\
     $F_{\rm var}$ $\oplus$ Lag amplitude $\oplus$ $z$ $\oplus$ log($C$)    & L-BFGS-B & tanh & 1 & 249 & 0.9993 [0.3541] & 0.0100 [0.4290]  \\
    \hline
    \\
     \end{tabular}
\end{center}
\end{table*}

\begin{figure*}
\centerline{
\includegraphics*[width=0.45\textwidth]{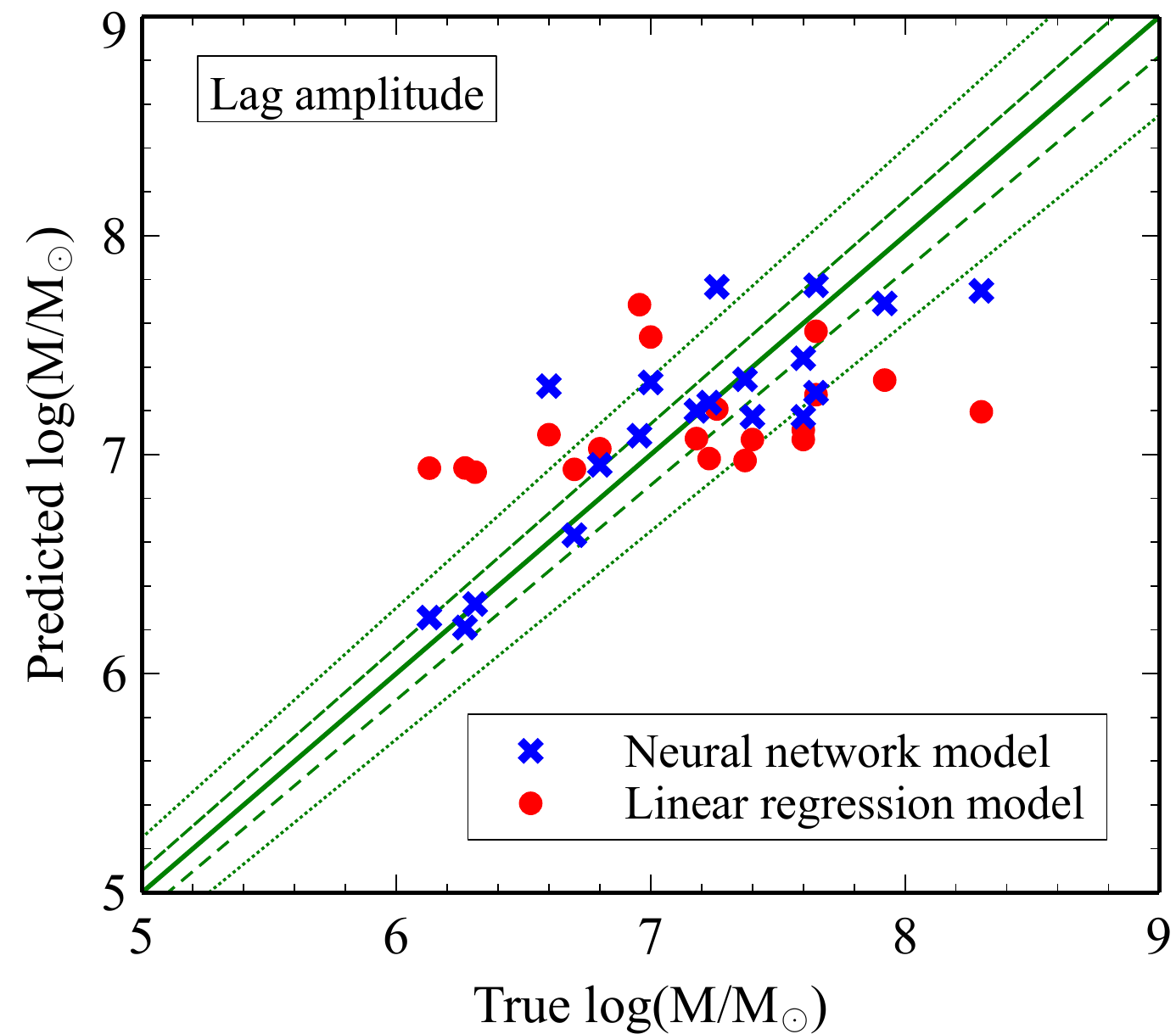}
\hspace{0.7cm}
\includegraphics*[width=0.45\textwidth]{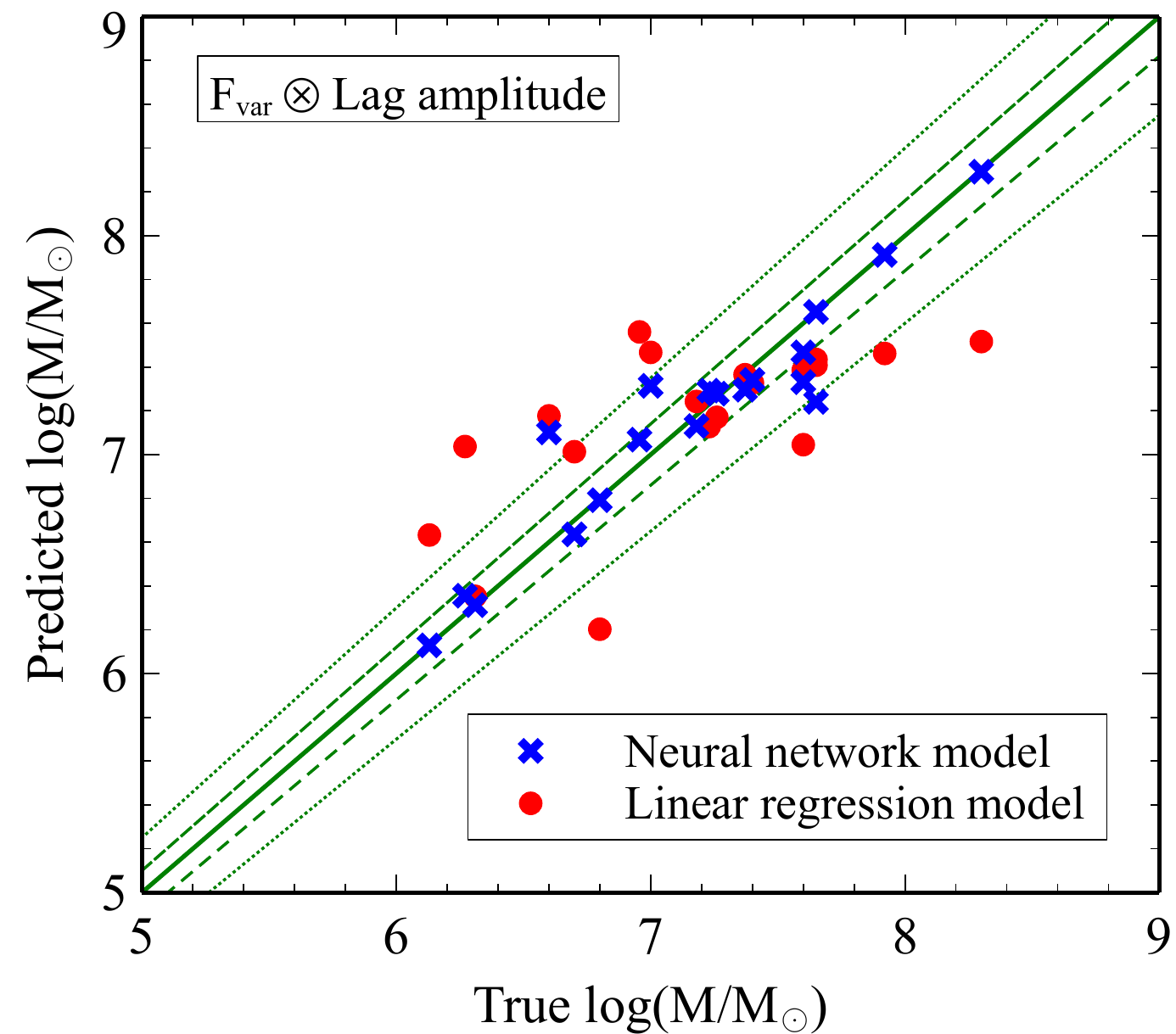}
\vspace{0.3cm}
}
\centerline{
\includegraphics[width=0.45\textwidth]{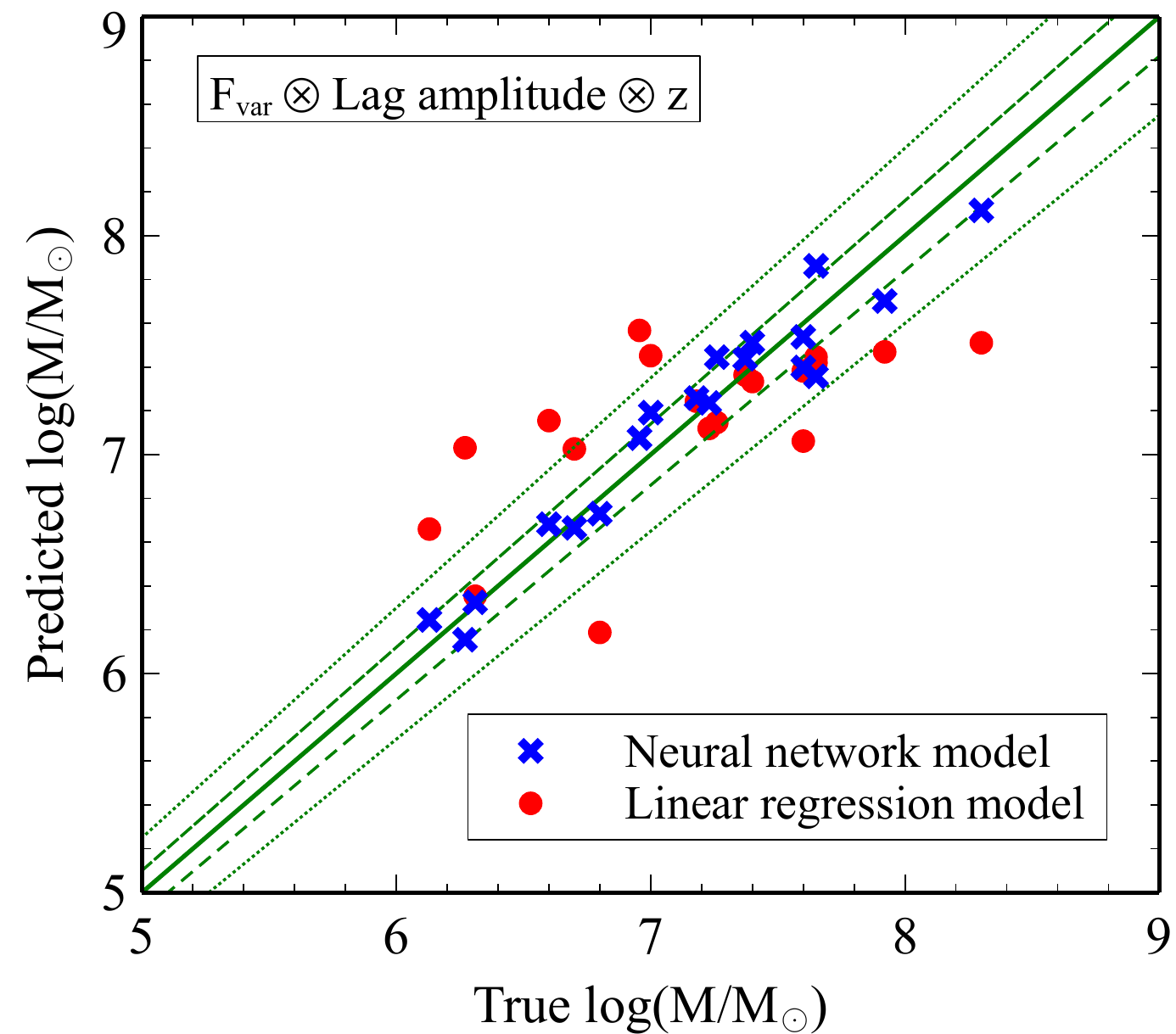}
\hspace{1.0cm}
\includegraphics[width=0.45\textwidth]{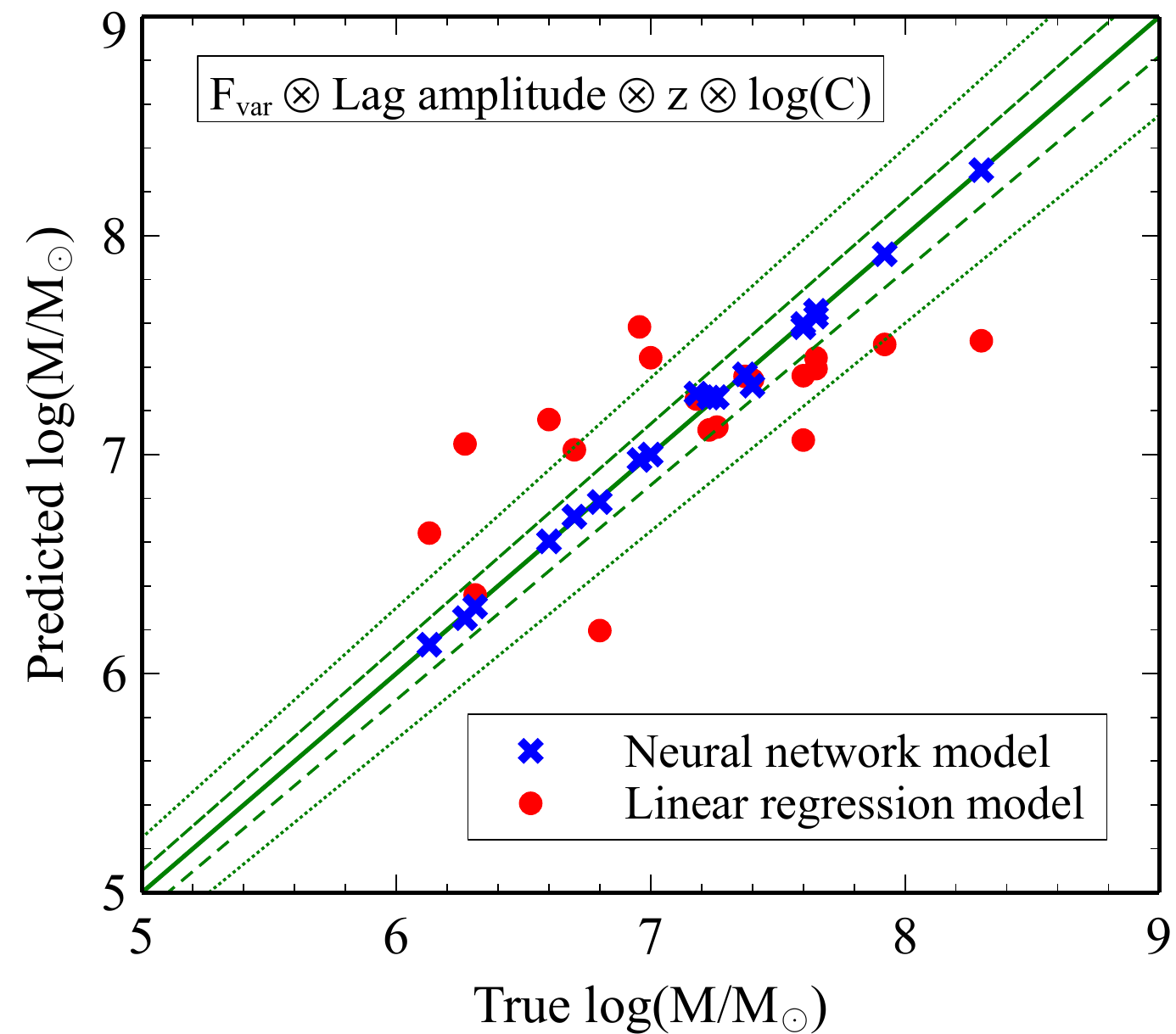}
}
\caption{Scatter plots of the true versus predicted values of the black hole mass. The blue-cross and red-circle data points represent the results obtained from the neural network and the linear regression model, respectively (Table~\ref{tab_ML_model}). The accuracy of the neural network model is significantly higher when using both the lag amplitude and $F_{\rm var}$ as the features of the model, and when compared with those of the linear regression model. The green solid lines show the perfect prediction lines. The region between two dashed lines and two dotted lines represents the parameter space where the deviations of the predicted black hole mass are still within 2 and 5 per cent from the true values, respectively. See text for more details.}
\label{fig_true_vs_pred}
\end{figure*}

Fig.~\ref{fig_true_vs_pred} shows the scatter plots of the actual values of the black hole mass and those predicted by the neural network models presented in Table~\ref{tab_ML_model}. For comparison, we also plot the results obtained when using the linear regression model. The linear regression model seems to overestimate the mass when $M\lesssim 10^{7}M_{\odot}$, but underestimate the mass when $M \gtrsim 10^{7}M_{\odot}$ (red data points in Fig.~\ref{fig_true_vs_pred}). Increasing the number of features does not improve the accuracy of the linear regression model. On the other hand, the predicted masses from the neural network model scatter around the true values with smaller errors when more features are included (blue data points in Fig.~\ref{fig_true_vs_pred}). We also draw the lines representing the region in the parameter space that covers 2 and 5 per cent of the mass deviation from the true values. It can be seen that the predicted mass using the combined features of the $F_{\rm var}$ and the lag amplitude can be constrained approximately within $\pm 5$ per cent of the true values. Using all features, we can place a constraint on the predicted mass to be $\pm 2$ per cent of the true mass.

All combined-feature models show $R^2 > 0.9$ even with 1 layer, so we do not increase the number of layers since it may result in unnecessary extreme time consuming while their accuracy cannot be significantly improved. We also investigate the case when the data of each feature are scaled to be between 0 and 1. The results are not much different to the cases when we do not scale them. The scaling factor then has a negligible effect on the efficiency of the model. 

\subsection{Lag and no-lag in reverberating and non-reverberating AGN}   

\begin{figure}
    \centerline{
        \includegraphics[width=0.45\textwidth]{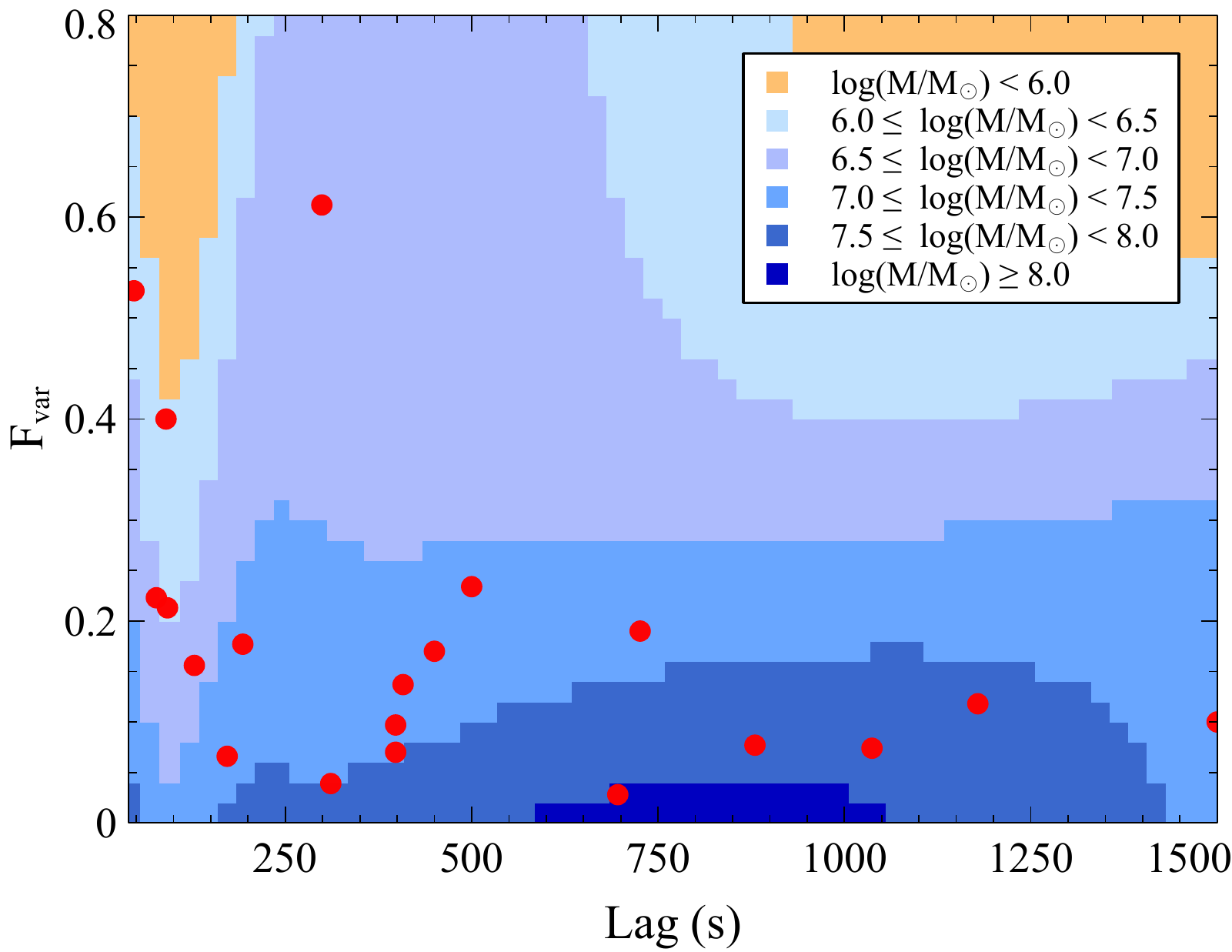}
    }
    \caption{Black hole masses plotted in the parameter space of our neural network model that employs both $F_{\rm var}$ and lag as the features. The observed reverberating AGN data are presented using red dots in this parameter space. Clearly, they occupy only the regime that implies the anti-correlation between the $F_{\rm var}$ and the lag. The Spearman’s rank correlation coefficients between the $F_{\rm var}$ and the lag of these observed reverberating AGN so far is $r_{s} = -0.46$.}
    \label{fig-para-space1}
\end{figure}

Furthermore, the $F_{\rm var}$--lag--mass model inferred from the neural network can be extrapolated to reveal the parameter space beyond what is occupied by the already-known reverberating AGN. We then plot the parameter distribution predicted by the model that includes both $F_{\rm var}$ and lag as the features in Fig.~\ref{fig-para-space1}, with the scattered red dots representing the current data of the observed reverberating AGN. It is clear that the AGN displaying a large $F_{\rm var}$ of $\sim 0.4-0.8$ while showing the lags of $\gtrsim 1000$ are not yet observed (i.e. top-right region of Fig.~\ref{fig-para-space1}). Our model predicts that the reverberating AGN belonging to this region, if exist, will have a small mass of $\lesssim 10^{6.5} M_{\odot}$, and the lag-mass scaling relationship may be weaker. There is, however, no simple explanation of how a small mass can induce such a large amplitude of the lag while still maintaining large $F_{\rm var}$. Perhaps, it is unphysical and that we still observe none of these reverberating AGN.

\begin{figure}
    \centerline{
        \includegraphics[width=0.45\textwidth]{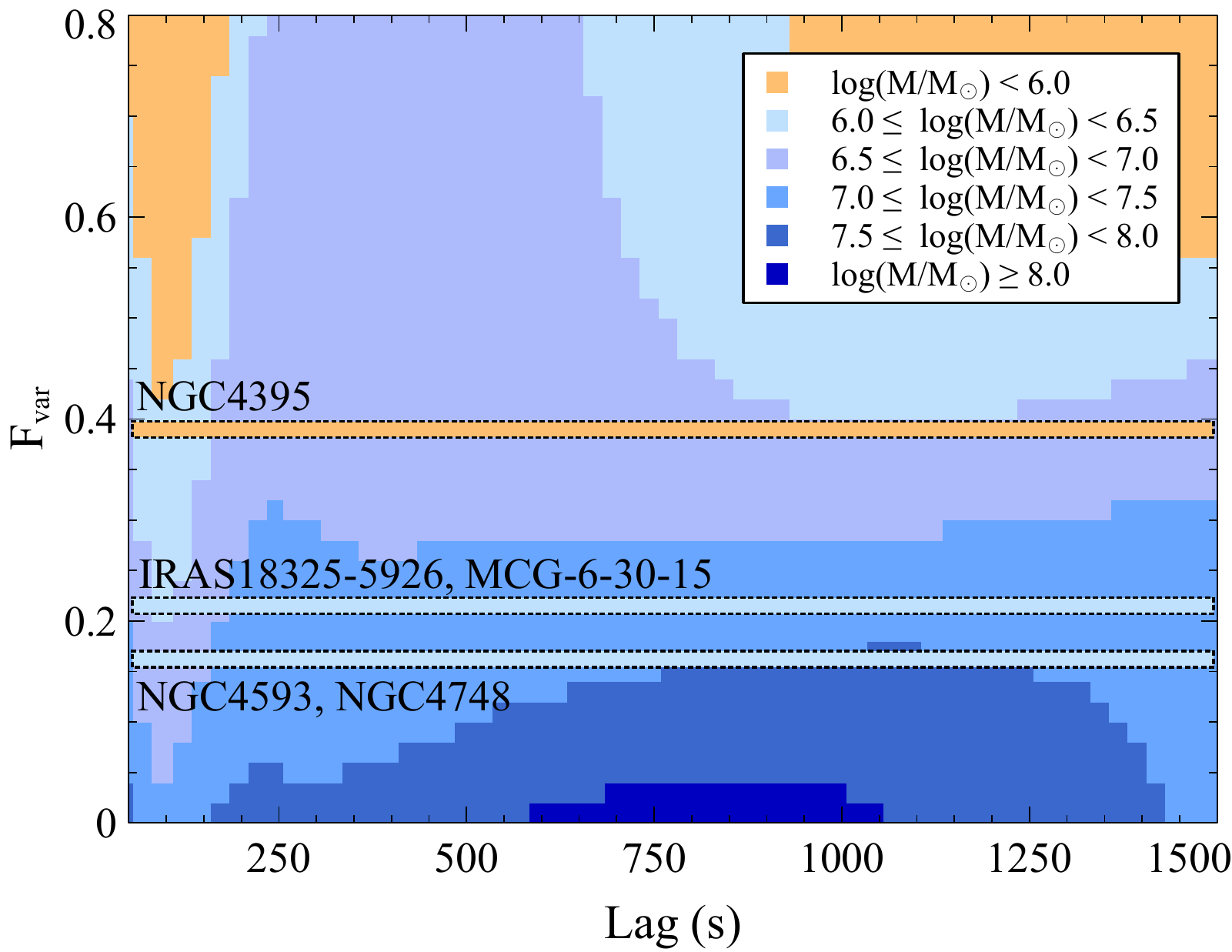}
    }
    \vspace{0.2cm}
    \centerline{
        \includegraphics[width=0.45\textwidth]{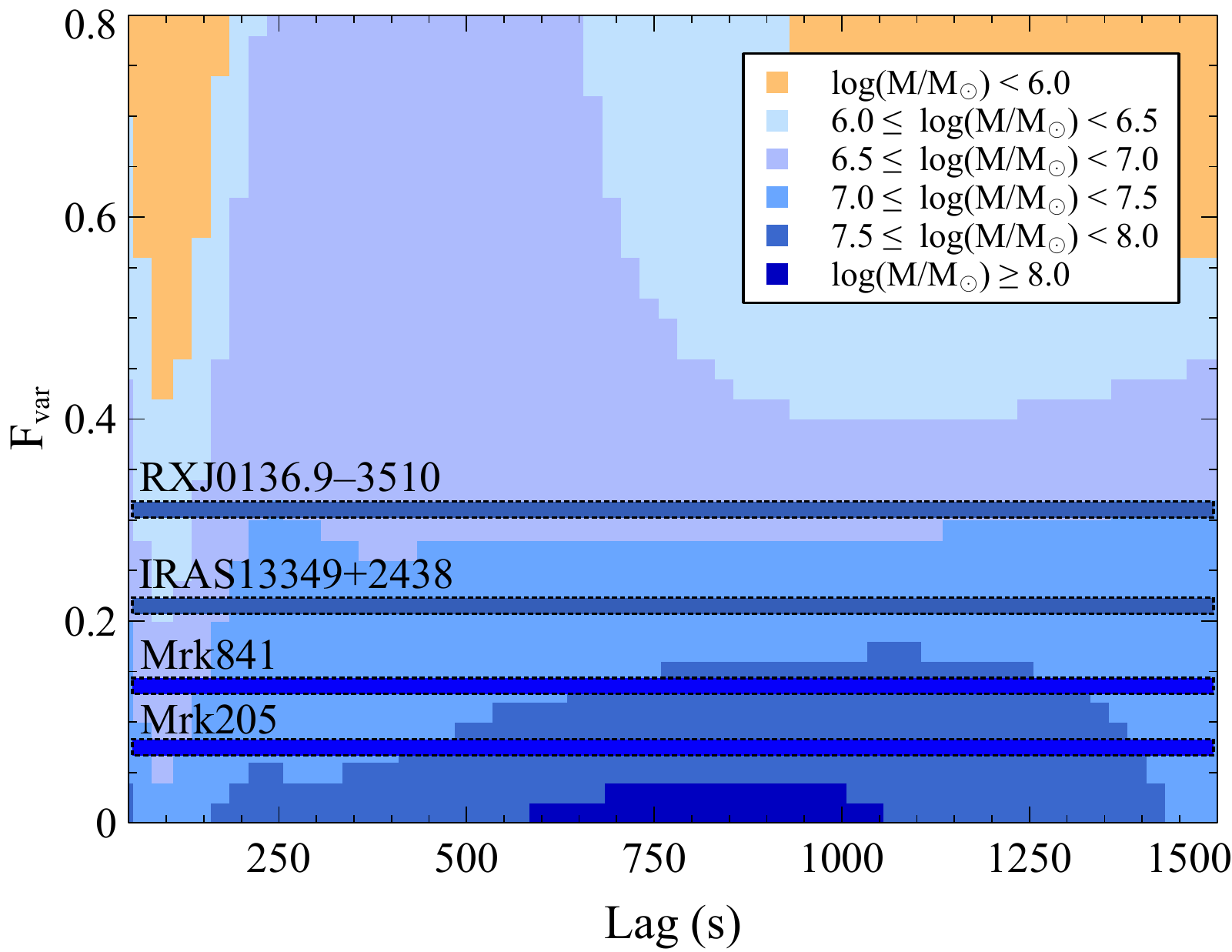}
    }
    \caption{Parameter distribution predicted by the model, as in Fig.~\ref{fig-para-space1}, over-plotted by the data of non-reverberating AGN. Their central mass and $F_{\rm var}$ are too small (top panel) and too large (bottom panel) that cannot fit into the constrained parameter space for the reverberating AGN, suggesting no observed Fe-K lag.}
    \label{non-reverb}
\end{figure}

Then, we apply the model to the AGN listed in Table~\ref{tab_predicted_M}. These AGN are completely new to the machine (i.e. kept unseen all the time during the training phase). They are non-reverberating AGN, however, so we use their mass and $F_{\rm var}$ to trace back to the lags that probably exist. We find that the model strongly suggests no observed lags for most of these sources, in agreement with the observations. The mass distribution of the non-reverberating AGN in the model parameter space is also illustrated as an example in Fig.~\ref{non-reverb}. It is clear that some have too small mass and $F_{\rm var}$ (e.g. ESO~113--G010, IRAS~18325--5926, MCG--6--30--15, NGC~4395, NGC~4593 and NGC~4748) that cannot fit into the constrained parameter space for the X-ray reverberating AGN. Some do not exhibit the lags because the central mass is large and the $F_{\rm var}$ is too large, against the strong anti-correlation of the mass and $F_{\rm var}$ predicted by the model (e.g. IRAS~13349+2438, Mrk~586, Mrk~704, PKS~0558--504 and RXJ~0136.9–-3510). 

Since the number of the reverberating AGN discovered so far is quite small, the ML prediction against the non-reverberating AGN samples can provide an indirect way to test the efficiency of the model. Regarding the high $R^2$ above 0.9 and the fact that the ML model can rule out the possibility to exhibit the lags in the majority of these non-reverberating AGN which are kept unseen during the training phase, we can ensure that the constrained $F_{\rm var}$--lag--mass relation (e.g. Fig.~\ref{fig-para-space1}) is reliable and that the model does not overfit the data.

\begin{table*}
\begin{center}
 \caption{Observed AGN data for final evaluation of the model. These are non-reverberating AGN that do not present clear Fe-K reverberation lags while showing variability that fits the criteria. The numbers in brackets denote the references where: (1) \protect\cite{Ponti2012}; (2) \protect\cite{Gonzalez2012}; (3) \protect\cite{Iwasawa2016}; (4) \protect\cite{Papadakis2010}. (R) indicates the optical reverberation mass estimate. The predicted properties of their time lags using the neural network model are presented in the final column, where `--' indicates no lags that is because either the mass is too low (`LM') or too high (`HM') that contradicts the constrained $F_{\rm var}$--lag--mass relation. The upper limits of the lags are reported if suggested by the model.} 
 \label{tab_predicted_M}
\begin{tabular}{lccc}
\hline
AGN name & true log($M/M_{\odot}$)  & $F_{\rm var}$ & Predicted lag properties \\
\hline
ESO 113--G010 & 6.74 (1) & 0.159 & --, LM  \\
ESO 511--G030 & 8.66 (1) & 0.050 & --, HM \\
IRAS 05078+1626 & 7.55 (1) & 0.063 & $\lesssim$ 1,300 s \\
IRAS 13349+2438 & 7.7 (2) & 0.211 &  --, HM \\
IRAS 18325--5926 & 6.4 (3) & 0.215 & --, LM \\
IZw1 & 7.4 (2) & 0.150 &  $\lesssim 600$ s  \\
MCG--02--14--009 & 7.13 (1) & 0.126 &  $\lesssim 500$ s \\
MCG--6--30--15 & 6.3 (1) & 0.212 & --, LM  \\
Mrk 1040	 & 7.6 (2) & 0.081 & $\lesssim$ 1,300 s  \\
Mrk 205	 & 8.32 (1) & 0.075 & --, HM \\
Mrk 586	 & 7.6 (2) & 0.248 & --, HM \\
Mrk 704	 & 8.11 (1) & 0.248 & --, HM \\
Mrk 766	 & 6.822 (R) & 0.228 & --, LM \\
Mrk 841	 & 8.52 (1) & 0.142 & --, HM \\ 
NGC 3227	 & 6.775 (R) & 0.096 & --, LM \\
NGC 3516	 & 7.395 (R) & 0.088 & $\lesssim 300$ s \\
NGC 4395	 & 5.449 (R) & 0.392 & --, LM \\
NGC 4593	 & 6.882 (R) & 0.172 & --, LM \\
NGC 4748	 & 6.407 (R) & 0.161 & --, LM \\
PKS 0558--504    & 7.8 (4) & 0.154 & --, HM \\
RXJ 0136.9--3510	 & 7.9 (2) & 0.315 & --, HM \\
\hline
\end{tabular}
\end{center}
\end{table*}
\nopagebreak

\subsection{Simulated new reverberating samples and parameter correlations}
Based on the $F_{\rm var}$--lag--mass parameter space constrained by the neural network model, we simulate 3200 new reverberating AGN samples to evaluate the obtained correlations. We investigate two possible cases: 1) when newly-discovered reverberating AGN samples are uniformly distributed into the constrained parameter space and 2) when newly-discovered reverberating AGN follow the current trend of the observed $F_{\rm var}$--lag anti-correlation. 

\begin{figure}
    \centerline{
        \includegraphics[width=0.45\textwidth]{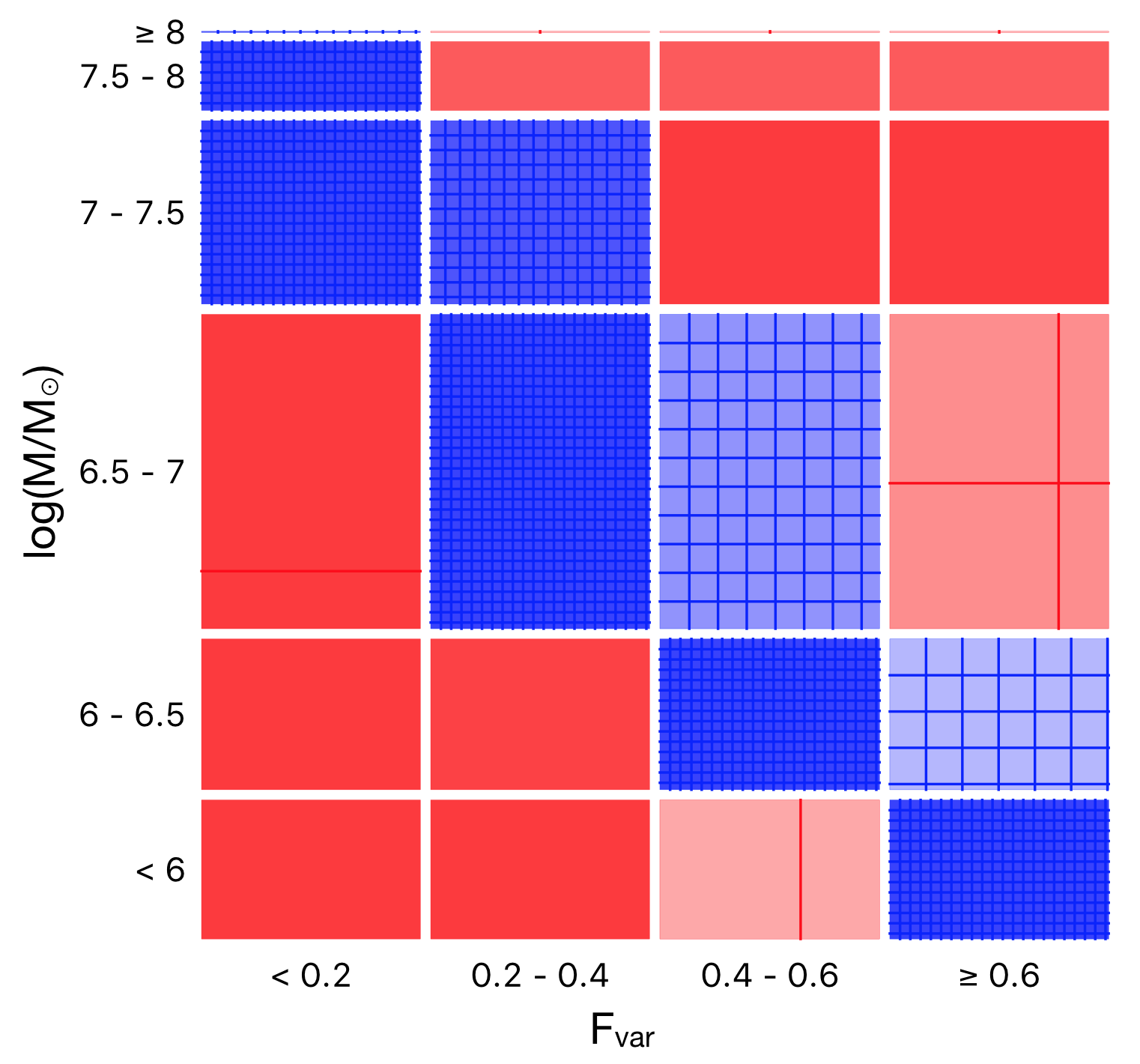}
    }
     \vspace{0.2cm}
    \centerline{
        \includegraphics[width=0.45\textwidth]{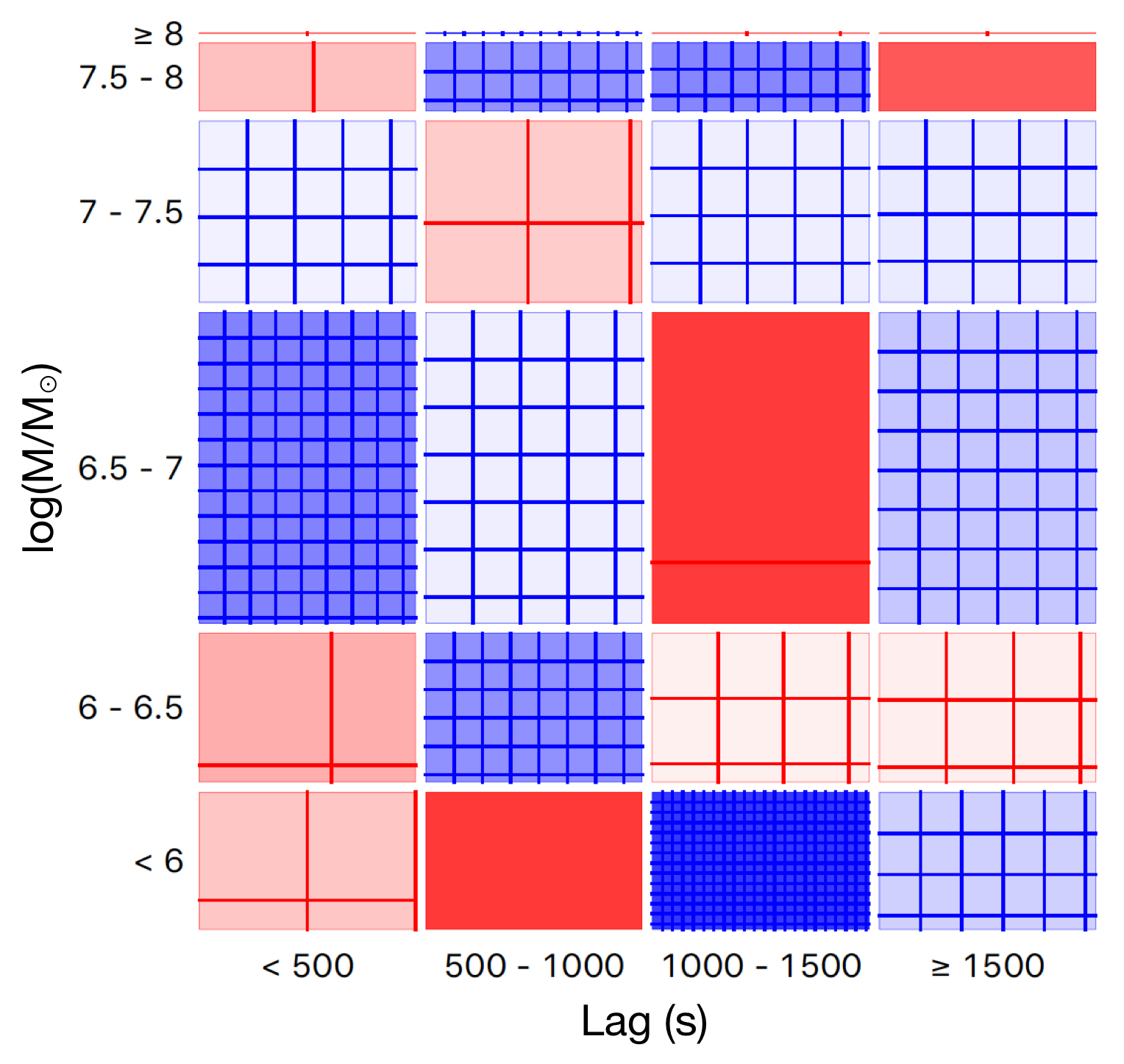}
    }
    \caption{Sieve diagrams of the mass and $F_{\rm var}$ (top panel) and the mass and lags (bottom panels) of the 3200 simulated reverberating AGN assuming a uniform distribution of the samples. The samples are drawn from the equally model-grid spacing of 0.02 for the $F_{\rm var}$ and 25~s for the lags (i.e. reverberating AGN has no preferable values of $F_{\rm var}$ and lags). Note that the expected frequency under independence and the observed frequency are proportional to the area of the rectangular and the corresponding number of inside squares, respectively. The differences between the observed and expected frequency are shown by the shading density. Blue (red) means the observed frequency is more (less) than the expected frequency. Under this assumption, $F_{\rm var}$--mass anti-correlation is stronger ($r_{s} = -0.88$) while the lag--mas correlation is drastically weaker ($r_{s} < 0.1$), compared to Fig.~\ref{fig-s}. See text for more details.}
    \label{fig-s2}
\end{figure}

Fig.~\ref{fig-s2} represents the Sieve diagrams for the black hole mass of the simulated 3200 reverberating AGN samples covering entire regime of the $F_{\rm var}$--lag parameter space. The model grids of $F_{\rm var}$ and lags with equal step sizes of 0.02 and 0.25~s, respectively, are used to produce a discrete uniform distribution of the samples. If the newly-discovered reverberating AGN lie uniformly in the constrained parameter regime, the $F_{\rm var}$ attribute associated with mass reveals a similar pattern as in Fig.~\ref{fig-s}, but with a higher correlation coefficient of $r_{s}=-0.883$. In the case of the lag, when associating it with mass, the lag--mass scaling relationship cannot be recovered. Intense blue rectangular at the lags of $\gtrsim 1000$~s and $M<10^{6}M_{\odot}$ means the number of actual data respondents to the AGN that has small mass with large lag becomes significantly more than what is expected. Interestingly, these results suggest that the $F_{\rm var}$--mass anti-correlation is always conserved and probably stronger with additional samples. The lag--mass relation, on the other hand, may be weaker if the observed $F_{\rm var}$ and lags become more independent with an increasing number of newly-discovered AGN.

We then investigate the case when newly-discovered reverberating AGN follow the current observed $F_{\rm var}$--lag anti-correlation ($r_{s} = -0.46$). A probability density function is employed to generate random values of the lags and $F_{\rm var}$ by allowing small deviations (2, 5, 10 and 15 per cent) from the current trend of $F_{\rm var}$--lag anti-correlation. Once the lags and $F_{\rm var}$ are drawn, the associating mass predicted by the neural network model can be assigned. We produce 3200 AGN sources and find that the lag--mass correlation coefficient in this case varies between $0.45 \lesssim r_{s} \lesssim 0.59$, which can be smaller or slightly larger than the correlation coefficient for only 22 observed reverberating AGN samples. Meanwhile, the $F_{\rm var}$--mass anti-correlation for these 3200 sources varies between $-0.82 \lesssim r_{s} \lesssim -0.71$, which is comparable or stronger than that of the reverberation samples so far. 

\begin{figure}
    \centerline{
        \includegraphics[width=0.45\textwidth]{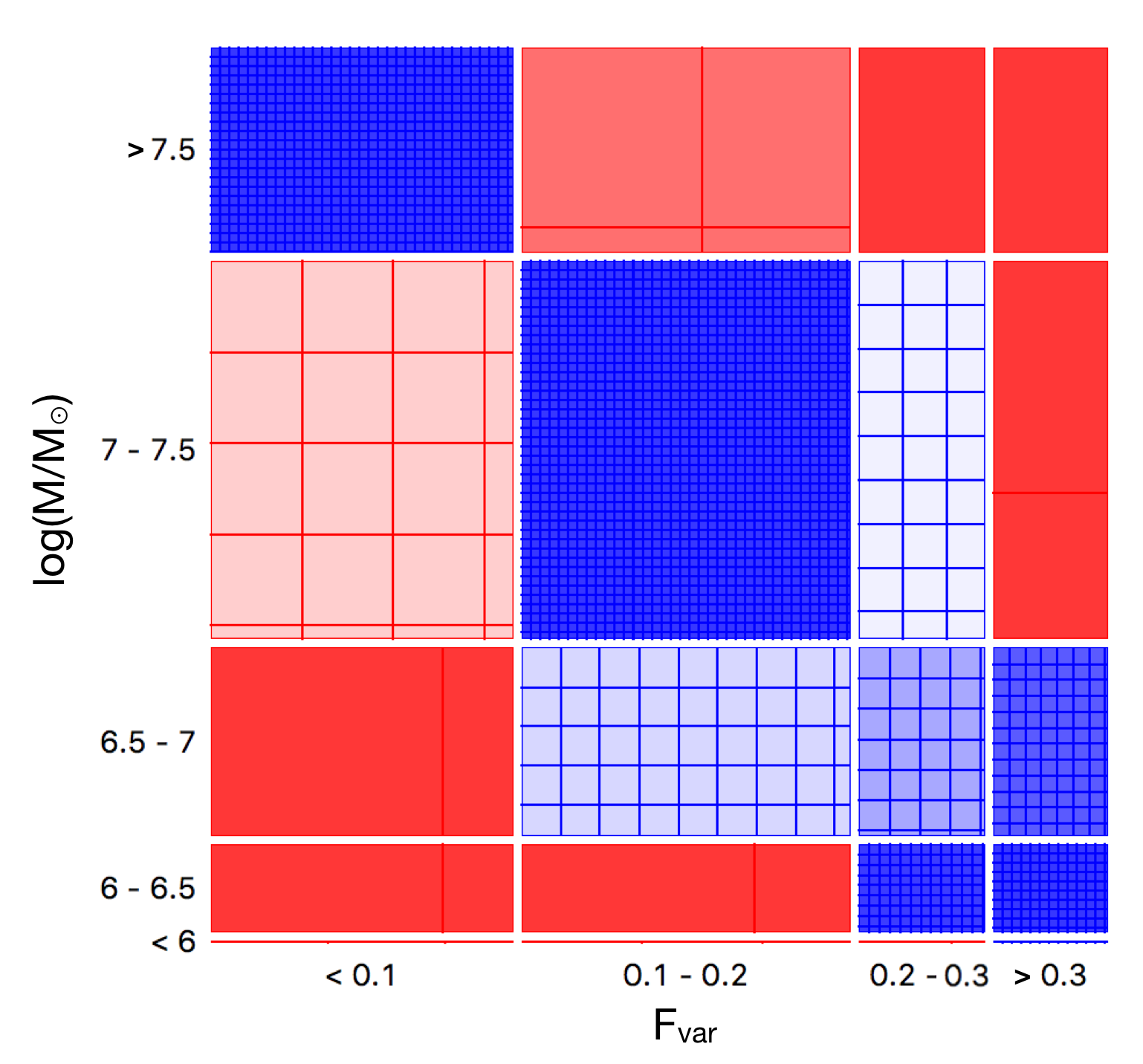}
    }
     \vspace{0.2cm}
    \centerline{
        \includegraphics[width=0.45\textwidth]{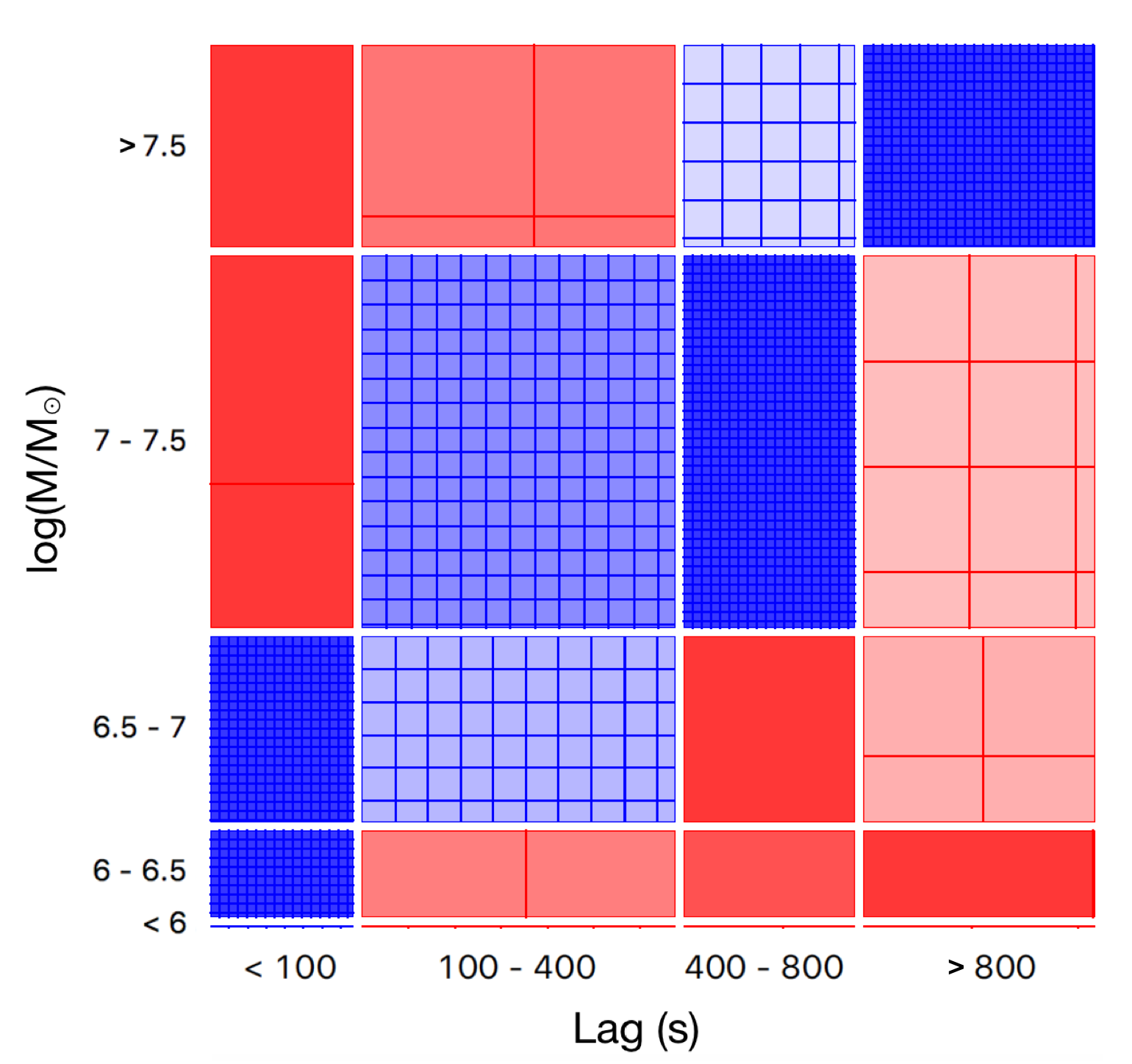}
    }
    \caption{Sieve diagrams of the mass and $F_{\rm var}$ (top panel) and the mass and lags (bottom panels) of 3200 reverberating AGN sources simulated under the constraint of the observed $F_{\rm var}$--lag relation ($r_{s} = -0.46$). In this illustration, the deviation of the $F_{\rm var}$--lag relation is $\sim 2$ per cent of the current trend. We find $r_{s}=0.59$ for the lags and the mass, and $r_{s} = -0.77$ for the $F_{\rm var}$ and the mass. These correlation coefficients are comparable to those of the observed 22 reverberating AGN (Fig.~\ref{fig-s}). Therefore, the lag--mass scaling relation is maintained through the tight anti-correlation between the lag and $F_{\rm var}$, and can be weaker if newly-discovered AGN samples are more uniformly distributed (Fig.~\ref{fig-s2}).}
    \label{fig-s2-2}
\end{figure}

\begin{figure}
    \centerline{
        \includegraphics[width=0.50\textwidth]{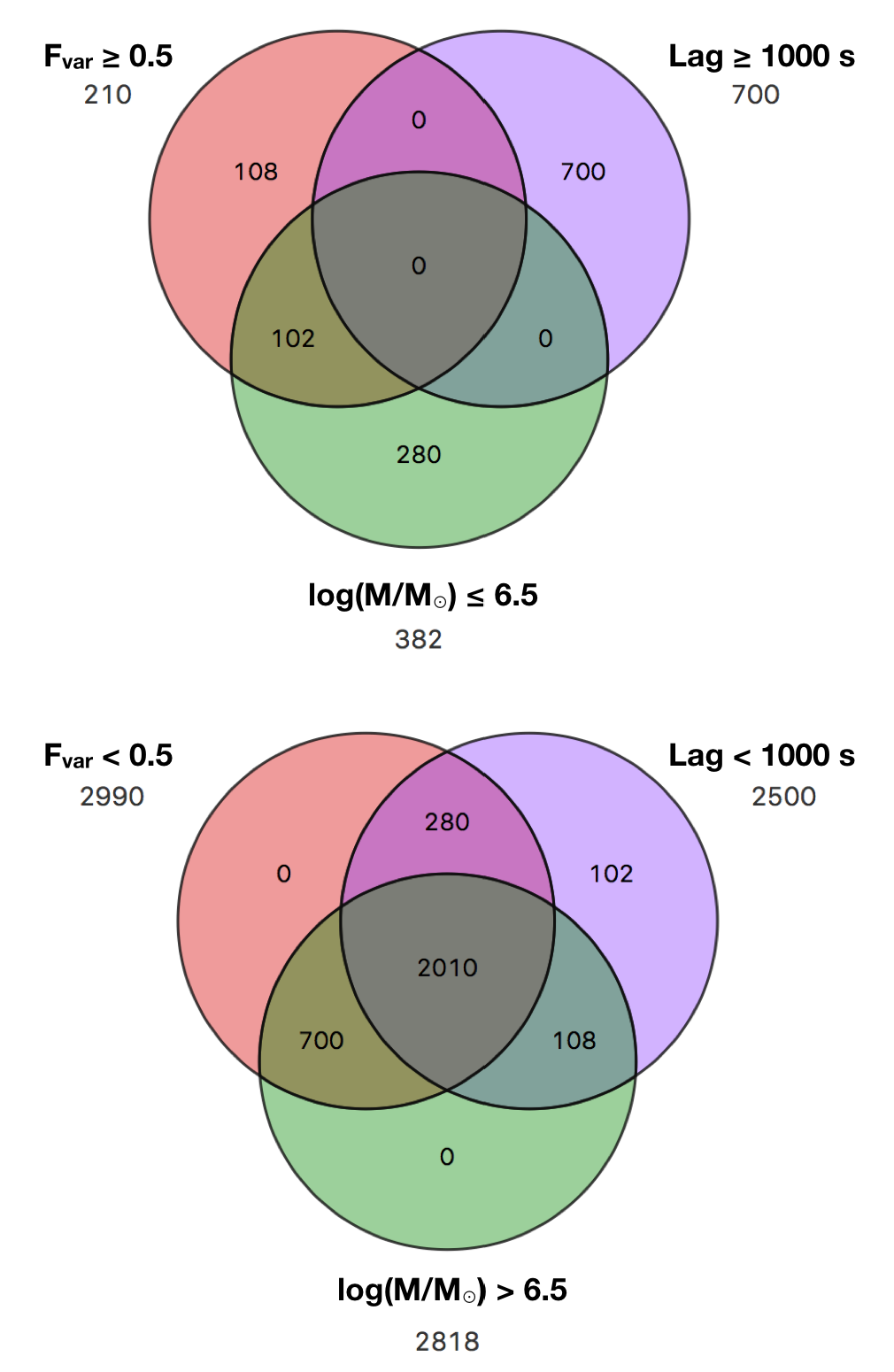}
    }
    \caption{Venn diagram showing the overlap of data instances from a collection of 3200 reverberating AGN sources, as in Fig.~\ref{fig-s2-2}, when the $\pm 2$ per cent deviation of the $F_{\rm var}$--lag relations from the current trend is allowed. The number of samples corresponding to each criterion is presented. Since the $\pm 2$ per cent deviation is small, we see no simulated sample that shows $F_{\rm var} \gtrsim 0.5$ with the lags $\gtrsim 1000$~s and $M \lesssim 10^{6.5} M_{\odot}$ (i.e. the top-right portion of the plot in Fig.~\ref{fig-para-space1}). Majority of the simulated samples shows $F_{\rm var} < 0.5$ with the lags $< 1000$~s and $M > 10^{6.5} M_{\odot}$, which is close to the trend of the observed reverberating AGN so far.}
    \label{fig-venn}
\end{figure}

\begin{figure}
    \centerline{
        \includegraphics[width=0.50\textwidth]{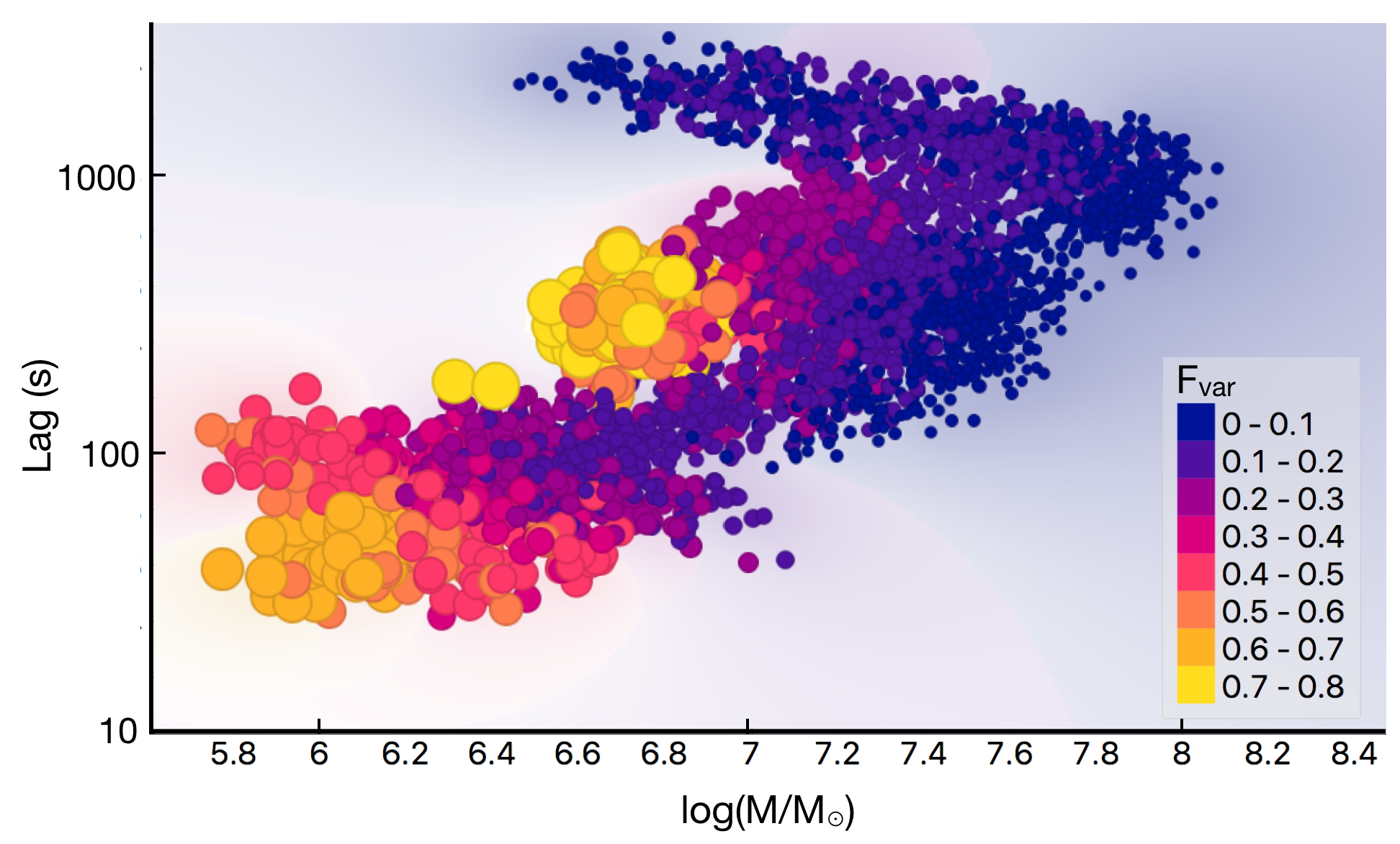}
    }
    \caption{Fe-K lag amplitude versus mass for 3200 newly-simulated reverberating AGN sources under the constraint of currently-observed $F_{\rm var}$--lag relations, with $\pm 2$ per cent deviations allowed. The size and colour of the data points relate to the values of the corresponding $F_{\rm var}$ (e.g. larger data-point size means larger $F_{\rm var}$). We find the Spearman's rank coefficient between the lag and the mass is +0.59, which is not significantly different from the obtained correlation coefficient for just 22 observed AGN samples in the {\it XMM-Newton} archive.}
    \label{fig-simulated-new-samples}
\end{figure}

Fig.~\ref{fig-s2-2} represents, as an example, the Sieve diagrams for the black hole mass of 3200 simulated AGN following the current trend of $F_{\rm var}$--lag relations, with $\pm 2$ per cent deviations allowed. The pattern of association as in Fig.~\ref{fig-s} is observed. The Venn diagrams for these newly-simulated AGN are presented in Fig.~\ref{fig-venn}. It can be seen that most of the simulated sources ($\sim 85$ per cent) are those that contain $M > 10^{6.5} M_{\odot}$ and show $F_{\rm var} < 0.5$, with $\sim 74$ per cent of these samples display the lags of $< 1000$~s. Note that this is an example in the case that we allow the deviation of the $F_{\rm var}$--lag relations from the current trend to be only $\pm 2$ per cent, so the samples associated with the top-right portion of the plot in Fig.~\ref{fig-para-space1} ($F_{\rm var} \gtrsim 0.5$, lags $\gtrsim 1000$~s and $M \lesssim 10^{6.5} M_{\odot}$) are not produced. Therefore, the lag--mass correlation coefficient does not change much. The corresponding Fe-K lag--mass relation is also presented in Fig.~\ref{fig-simulated-new-samples}. These results suggest that the tight anti-correlation between the lag and $F_{\rm var}$ is necessary to maintain the lag--mass scaling relation. On the other hand, the $F_{\rm var}$--mass anti-correlation is likely stronger whether or not new reverberating AGN show dependence of their lags on $F_{\rm var}$. 

\section{Discussion}

According to our results, the developed ML models can make higher accurate predictions of the black hole mass than the linear regression model. The combination of the solver and activation function is dependent on the characteristics of the data. All neural network models here require L-BFGS-B solver. Generally, L-BFGS-B solver is a suitable second-order optimization method and is fastest for small convex problems or small data size, appropriate for our data set. It employs the full training set to obtain the later update to parameters at every iteration \citep{Le2011}. If the size of the dataset is large, L-BFGS-B computing time can be very long on a single machine, so it is hard to incorporate new data in an online environment. On the other hand, tanh is found to be the best activation function that provides high accuracy for generalized MLP architectures of neural networks \citep{Karlik2011, Montavon2012}. If the $F_{\rm var}$ is only a single feature, the correlation between $F_{\rm var}$ and log($M/M_{\odot}$) is linearly dependent. This is why the ReLu which is rectified linear unit activation functions is  better than tanh (full non-linearity unit). All combined-feature models ($R^{2} \gtrsim 0.9$) developed here require the L-BFGS-B solver and tanh activation function, which agree with other works that use multi-feature/variable regression in MLP neural networks \citep[e.g.][]{Gheorghe2014, Artrith2017}.

\cite{Kara2016} found that the correlation coefficient of the frequency–mass relation ($r_{s}=-0.68$) is stronger than that of the lag–mass relation ($r_{s}=0.60$). They suggested using the frequency–mass relation for determining the black hole mass instead of the lag–mass relation. This argument, however, may be specific to the averaging scheme applied to the observational data (Hancock et al., in prep.). Here, we find the correlation between the Fe-K lag amplitude and the mass for AGN sources to be $r_{s}=0.589$, confirming what was reported by \cite{Kara2016}. Even though both $F_{\rm var}$ and log($C$) are produced from the lightcurves extracted in the same energy band of 2--10 keV, it is clear that the black hole mass shows a significantly stronger correlation with $F_{\rm var}$ than log($C$). Also, both log($L_{\rm bol}$) and log($C$) show a relatively weak correlation with the black hole mass. This suggests that the X-ray timing data may be more useful for predicting the black hole mass than the time-average X-ray data.

Hancock et al. (in prep.) studied the correlations between the central mass and the lags, but using the lags measured between the soft excess 0.3--0.8 keV and continuum-dominated 1--4 keV bands. The correlation coefficient was found to be $r_{s}=0.72$. This may suggest that the lags in the soft band are stronger correlated to the mass than the lags in Fe-K band. Although the Fe-K band is sometimes contaminated by the neutral distant reflection, the distant reflection varies on different timescales from the timescales of the inner-disc reflection. Therefore, unlike the soft excess lags whose origin is ambiguous, the Fe-K lags are likely to be a clean signature of X-ray reverberation. The Fe-K band then may provide a more accurate reflection of how much the signals produced via X-ray reverberation correlate with the black hole mass.  

We also analyze the correlation between the black hole mass and other features of the ML model. Interestingly, we find that the correlation between the mass and the $F_{\rm var}$ is $r_{s}=-0.718$, which is stronger than the lag--mass relation and even stronger than the frequency--mass relation reported by \cite{Kara2016}. However, the prediction accuracy is higher when using the lags alone than when using the $F_{\rm var}$ alone. Nevertheless, for a specific source, different individual or combined observations into, e.g., high flux and low flux states result in different time lag estimates for a single mass. Therefore, only one parameter cannot be a good predictor for the mass since its value changes with how the data of each AGN are selected and combined while the mass remains constant. It is then better to use both $F_{\rm var}$ and lag amplitude as the mass predictors that, evidently, can also improve the model accuracy ($R^{2} = 0.9124$). 

Based on the $F_{\rm var}$--lag--mass relation predicted by the ML model, the reverberating AGN with a specific black hole mass can exhibit $F_{\rm var}$ and lags within a limited regime. The model can rule out the possibility for the majority of the non-reverberating AGN to display the lags (Table~\ref{tab_predicted_M} and Fig.~\ref{non-reverb}), revealing its potential to make accurate predictions for new data. Since the model can be successfully applied to both reverberating and non-reverberating AGN, it suggests the common origins of main variability that contributes to $F_{\rm var}$ for both AGN groups. This suits the framework of the X-ray variability driven by, e.g, the disc propagating-fluctuations that can operate in both reverberating and non-reverberating AGN.

Note that the model still predicts the presence of the Fe-K lags for some sources included in the non-reverberation group (Table~\ref{tab_predicted_M}). For example, the Fe-K lags of IZw1 are predicted to be $\lesssim 600$~s. In fact, this is in agreement with \cite{Wilkins2021} who reported observations of X-ray flares around the central black hole in IZw1 and detected the Fe-K reverberation lags of $\sim 746 \pm 157$~s by using a different method to the standard Fourier analysis. Moreover, based on the \emph{XMM-Newton} observations, \cite{Zoghbi2013} reported the Fe-K reverberation lags in MCG--5--23--16 and SWIFT~J2127.4+5654, but new analysis using \emph{NuSTAR} data suggested no strong evidence for relativistic reverberation in both AGN \citep{Zoghbi2021}. It is not straightforward to determine if these AGN exhibit no intrinsic reverberation lags or if the inconsistencies arise due to different methods and model assumptions used to quantify the significance of the lag. Nevertheless, we try moving these samples to the non-reverberating AGN group, and find only a slight change in the lag--mass correlations inferred (e.g., lag-mass correlation coefficient $r_{s}$ changes less than $\sim 4$ per cent). This is because these inconsistencies are found only in the minority of the samples. Therefore, moving them across different groups (reverberation and non-reverberation) does not change the trend of the key results here.

The X-ray reverberation produces the dip in the PSD profiles so it dilutes the $F_{\rm var}$ in a particular timescale where reverberation dominates \citep{Papadakis2016, Chainakun2019a}. The amplitude of the dip increases with increasing the reflection fraction, resulting in a decrease in $F_{\rm var}$ on a particular reverberation timescales. The mass, on the other hand, scales up the intrinsic lags. However, $F_{\rm var}$ here represents the fractional excess variance contributed by all mechanisms that affect the X-ray variability in 2--10 keV band, not only by the reverberation. For example, the $F_{\rm var}$ can be affected by the constant or less-variable emission component, e.g. from outflowing gas, that is varied among different AGN \citep{Parker2021}. The constant emission component raises the mean but does not affect the standard deviation of the data, so the $F_{\rm var}$ decreases. On the other hand, the constant absorption lowers both mean and standard deviation of the data proportionally, hence its presence has small effects on the $F_{\rm var}$. The deviation of the $F_{\rm var}$--lag relation then may be induced by, e.g., the presence of the constant or less-variable emission component, resulting in the change of the lag--mass correlation coefficient.  

Our results suggest that to maintain the lag--mass scaling relation, the anti-correlation between the lag and $F_{\rm var}$ must preserve. Contrarily, the $F_{\rm var}$--mass anti-correlation seems to be stronger even if new reverberating AGN show more independence of their lags on $F_{\rm var}$. Moreover, the model suggests the regime of $F_{\rm var} \gtrsim 0.5$, lags $\gtrsim 1000$~s and $M \lesssim 10^{6.5} M_{\odot}$ where the reverberating AGN are not yet observed (top-right portion of Fig.~\ref{fig-para-space1}). It possibly represents a forbidden regime that is unphysical, which is why there is still no reverberating AGN sample fitting into. Although the AGN with constant emission component can introduce the deviation of the $F_{\rm var}$--lag correlation coefficient, they may not fit into this ambiguous regime. This regime corresponds to the samples with high $F_{\rm var}$, while constant emission produces negative effects on $F_{\rm var}$.

Furthermore, the observed flux in a particular energy band always contains both continuum and reflection flux. This introduces dilution effects to the lags so that the observed lags are smaller than the intrinsic lags \cite[e.g.][]{Wilkins2013, Chainakun2015}. \cite{Chainakun2019b} studied the lags produced by an extended corona under the inverse-Compton scattering scenario and found that, with a fixed coronal size and geometry, the coronal temperature and optical depth can affect the lags. The more the complex model is, the less the measured reverberation lags can straightforwardly relate to the true light-travel distance. In fact, \cite{Hinkle2021} investigated the fundamental X-ray corona properties of 33 AGN observed under the 105-month \emph{Swift/BAT} campaign together with the archival \emph{XMM-Newton} and \emph{NuSTAR} data. They found no strong correlations between the black hole mass, coronal compactness and coronal temperature. Therefore, if the lags significantly depend on the coronal compactness or temperature, the lags may not strongly correlate with the mass. The less correlation between the lags and the mass, if observed, then may suggest the extended corona framework where the lag amplitudes of the majority of the sources are more strongly affected by the properties of the coronal extent rather than the geometry.

\begin{figure*}
    \centering
    \includegraphics*[scale=0.5]{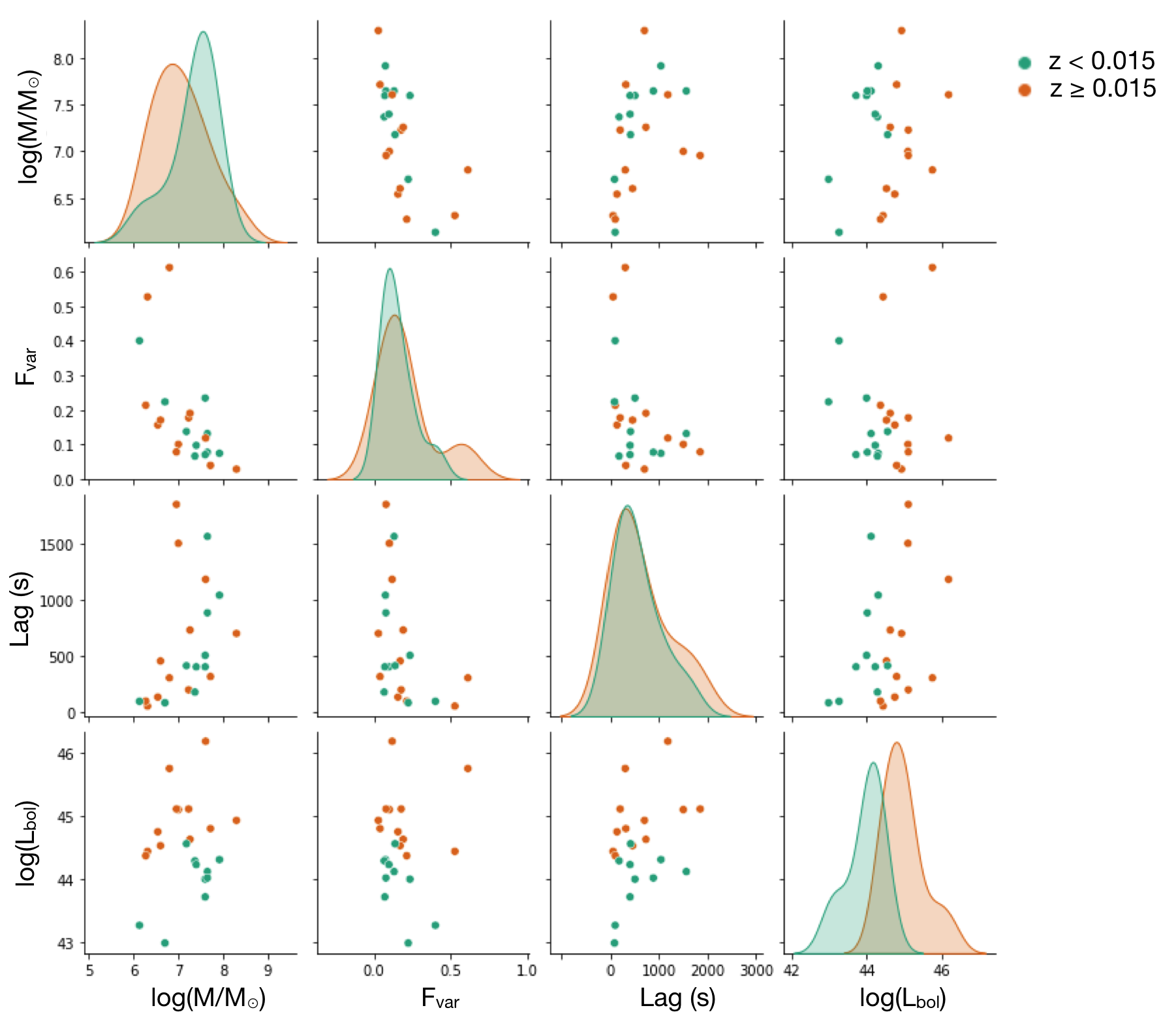}
    \caption{Pairwise relationships between the variables of current reverberating AGN samples when the redshift $z$ is discretized into 2 groups: $z < 0.015$ (low redshift) and $z \geq 0.015$ (high redshift). The diagonal plots represent a univariate marginal distribution of the data in each column, derived from a layered kernel density estimate (KDE). We can see the high redshift samples occupy the parameter space associated with high $L_{\rm bol}$. However, the clear separation between low and high $z$ samples cannot be clearly seen when the data are plotted in the lag, mass and $F_{\rm var}$ parameter space. The amount of current \emph{XMM-Newton} samples is not enough for the ML model to gain useful information of the lag, mass and $F_{\rm var}$ correlating with $z$. 
     }
    \label{fig-12}
\end{figure*}

According to our results, the $F_{\rm var}$ and lag amplitude are the key parameters of the X-ray variability data to predict the black hole mass. The clear trend of how the mass varies with log($C$) and $z$ is not clearly seen. Based on the current \emph{XMM-Newton} observations, we find $z$ is correlated with $L_{\rm bol}$ ($r_{s}=0.796$), but there is no clear pattern of associations between z and other parameters (time lags, mass, and $F_{\rm var}$). Perhaps, this is because the samples probed by \emph{XMM-Newton} are only those relatively nearby and are not distributed widely enough across a broad range of redshift. Fig.~\ref{fig-12} represents the distribution of our reverberating AGN samples into the parameter space when the redshift $z$ is discretized into 2 groups: $z < 0.015$ (low redshift) and $z \geq 0.015$ (high redshift). The results show that we tend to observe the high-redshift sources when they exhibit relatively large $L_{\rm bol}$. The pattern of associations between $z$ and other parameters cannot be easily revealed, however, even though the samples are just grouped into low and high redshifts. This is why we still cannot obtain a meaningful interpretation for the redshift associated with the lag and the mass. Note that these results do not imply that there is no effect of redshift at all, they instead suggest that the amount of current samples is not enough for the model to gain useful information relating to $z$. In the \emph{Athena} era, the model should be able to provide more insights into not only the trend of scaling relations, but also the redshift dependence, when we can probe more reverberating sources at higher redshift. 

It is also possible to calculate $F_{\rm var}$ in the frequency domain by integrating the corresponding power spectrum between two frequencies, so that the obtained $F_{\rm var}$ is specific to a particular range of timescales, which will be investigated in the future work. Finally, there were also reports on the correlation between the central black hole mass and the host galaxy total stellar mass \citep[e.g.][]{Reines2015}. Investigating the relationship between the AGN and host-galaxy parameters using machine learning techniques is also planned for the future. 

\section{Conclusion}

We investigate several scenarios of applying the neural network to predict the black hole mass and correlations in AGN. The model does not require the assumed source geometry in advance, but this does not mean that the inferred correlations are independent of source geometry. The predictions by the neural network models are much more accurate than those from the standard linear regression in all cases, providing an independent way to predict the AGN mass. The best model that uses either $F_{\rm var}$ or the lag alone contains $\sim 90$--130 neurons ($R^{2}\sim 0.6$--0.7 and MAE $\sim0.2$--0.3). It is, however, better to use both $F_{\rm var}$ and the lag amplitude to predict the central mass ($R^{2}=0.9124$ and MAE = 0.1077), which requires 168 neurons. The mass distribution predicted by the model satisfies the lag--mass scaling relation regarding to the observed reverberating AGN samples. Despite of the limited number of available samples for training the machine, we illustrate that the model can potentially be used to exclude and identify the AGN samples that do not exhibit the lags. This, in turn, suggests the common origins of the variability, e.g. disc propagating-fluctuations, that drives $F_{\rm var}$ in both reverberating and non-reverberating AGN. 
 
The $F_{\rm var}$--mass anti-correlation is always true and probably stronger with an increasing number of sources. The model reveals the regime in the parameter space where there is still no reverberating AGN samples lying into ($F_{\rm var} \gtrsim 0.5$ with the lags $\gtrsim 1000$~s and $M \lesssim 10^{6.5} M_{\odot}$). In fact, there is no straightforward mechanism for the AGN to operate high $F_{\rm var}$ and large lags with such a small mass. Perhaps, this is the reason why we do not observe reverberating AGN in this regime. The model shows that the less correlation between the lag and the mass is possible with increasing number of the newly-discovered reverberating AGN. The AGN inducing this may require an extended corona whose property (e.g. temperature and compactness) rather than geometry is more dominant to the measured time lags. The hints of their presence then may be a noticeable decrease in the lag–mass correlation coefficient.  
 
The more accurate mass can be retrieved using all features investigated here ($R^{2}=0.9993$ and MAE = 0.0100). The neural network in this case contains 249 neurons. The model with all features coupled together is probably too complex, so the clear pattern of association to the relation between $z$ and log($C$) is not observed. The obtained neural network models are observational driven since they are trained and tested using the real observational data. The current Fe-K reverberation samples observed so far are representative enough in the way that the model can draw some meaningful correlations between their observable variables. It is possible that the source spectral state may have an effect on the detection probability. For example, super soft sources occupying at the high end of the $L_{\rm bol}$ parameter space may have poor count rates in the hard band so detection of Fe-K lags become more difficult. Regarding the obtained correlation between $L_{\rm bol}$ and the mass which is very weak (Table~\ref{tab_corr}), it will require a very large number of these sources in order to drive the significant and meaningful relationship between $L_{\rm bol}$ and the mass in the samples. We suspect that these extreme sources are likely not the majority of populations, hence do not significantly alter the inferred correlations. This work illustrates an application of the ML technique to construct the multi-feature parameter space where the samples of reverberating AGN can be simulated. More number of newly-discovered reverberating AGN will help place robust constraint on the extrapolating results predicted by the models. 

\section*{Acknowledgements}
This research has received funding support from the NSRF via the Program Management Unit for Human Resources \& Institutional Development, Research and Innovation (grant number B16F640076). The ML models in this work are adopted from the {\sc sklearn} software package. The calculations were carried out using the high performance computing resources in the Institute of Science and the Centre for Scientific and Technological Equipment, Suranaree University of Technology.

\section*{Data availability}
The data underlying this article can be accessed from {\it XMM-Newton} Observatory (\url{http://nxsa.esac.esa.int}). The neural network algorithm adopted here is available in {\sc scikit-learn} at \url{https://scikit-learn.org/}. The derived data and developed model underlying this article will be shared on reasonable request to the corresponding author.


\bibliographystyle{mnras}

\bsp	
\label{lastpage}
\end{document}